\documentclass[12pt]{article}
\usepackage{amssymb}
\usepackage{amsmath}
\usepackage{amstext}
\usepackage{graphicx,epsfig}
\usepackage{epsfig}
\usepackage{verbatim} 
\usepackage{fancybox}
\usepackage{color}
\usepackage{enumitem}
\usepackage{subfigure}
\usepackage{bbm}
\usepackage{parskip}
\usepackage{dsfont}
\usepackage[numbers,sort&compress]{natbib}
\usepackage{verbatim}
\usepackage{appendix}

\newcommand{\D}{{\rm d}}

\newcommand{\Comment}[1]{{}}
\definecolor{darkblue}{rgb}{0.95,0.05,0.05}
\definecolor{reddish}{rgb}{0.95, 0.05, 0.05}
\usepackage[linktocpage=true]{hyperref}
\hypersetup{
colorlinks=true,
citecolor=darkblue,
linkcolor=reddish,
urlcolor=darkblue,
pdfauthor={},
pdftitle={},
pdfsubject={}
}

\newcommand{\nn}{\nonumber \\ }
\newcommand{\f}{\frac}

\def\({\left(}
\def\){\right)}

\newcommand{\beq}{\begin{equation}}
\newcommand{\eeq}{\end{equation}}
\newcommand{\be}{\begin{equation}}
\newcommand{\ee}{\end{equation}}
\newcommand{\bea}{\begin{align}}
\newcommand{\eea}{\end{align}}

\newcommand{\vp}{\varphi}

\newcommand{\la}{\langle}
\newcommand{\ra}{\rangle}

\def\gsim{ \lower .75ex \hbox{$\sim$} \llap{\raise .27ex \hbox{$>$}} }
\def\lsim{ \lower .75ex \hbox{$\sim$} \llap{\raise .27ex \hbox{$<$}} }

\def\g {{\gamma}}
\def\beq{\begin{eqnarray}}
\def\eeq{\end{eqnarray}}
\def\m{M_*}

\def\d{\mathrm{d}}
\def\p{{\cal P}}

\def\s{\sigma}
\def\g{{\cal G}}

\def\L*{{\cal L}_*}
\def\L{\mathcal{L}}
\def\({\left(}
\def\){\right)}

\def\nn{\nonumber}
\def\p{\partial}

\def\p{\partial}

\def\<{\langle}
\def\>{\rangle}

\def\xyma{\xymatrix@M.7em}
\def\xymas{\xymatrix@M.1em}
\newcommand{\ba}{\begin{eqnarray}}
\newcommand{\ea}{\end{eqnarray}}

\def\a{\alpha}
\def\b{\beta}
\def\c{\chi}

\def\d{\delta}
\def\D{\Delta}

\def\f{\frac}
\def\g{\gamma}

\def\l{\left}

\def\la{\langle}
\def\ra{\rangle}

\def\mc{\mathcal}

\def\m{\mu}
\def\n{\nu}

\def\nn{\nonumber}

\def\om{\omega}
\def\Om{\Omega}

\def\p{\partial}

\def\r{\right}
\def\s{\sigma}
\def\t{\theta}

\def\vp{\varphi}

\def\x{\xi}

\def\be{\begin{equation}}
\def\ee{\end{equation}}

\def\bea{\begin{eqnarray}}
\def\eea{\end{eqnarray}}

\def\ba{\begin{array}}
\def\ea{\end{array}}

\def\bc{\begin{center}}
\def\ec{\end{center}}

\def\bl{\begin{flushleft}}
\def\el{\end{flushleft}}

\def\br{\begin{flushright}}
\def\er{\end{flushright}}

\def\bi{\begin{itemize}}
\def\ei{\end{itemize}}

\def\bt{\begin{tabular}}
\def\et{\end{tabular}}

\def\wh{\widehat}

\title{}
\author{}

\numberwithin{equation}{section}

\begin{document}

\renewcommand{\thefootnote}{\fnsymbol{footnote}}

\begin{center}
\bf \Large{Semiclassics, Goldstone Bosons and CFT data}
\end{center}
 \vspace{1truecm}
\thispagestyle{empty} 

\begin{center}
{\textsc {A.~Monin$^{\dagger}$, D.~Pirtskhalava$^{\dagger}$, R.~Rattazzi$^{\dagger}$, F.K.~Seibold $^{\dagger\ddagger}$}}
\end{center}
\vspace{.5cm}

\centerline{{\it $^{\dagger}$ Institute of Physics,}}  \centerline{{\it \'Ecole Polytechnique F\'ed\'erale de Lausanne,}}
\centerline{{\it CH-1015, Lausanne, Switzerland}}

\vspace*{0.5cm}
\centerline{\it $^{\ddagger}$ Institute for Theoretical Physics,}
\centerline{\it Eidgenössische Technische Hochschule Zürich,}
\centerline{\it Wolfgang-Pauli-Strasse 27, 8093 Zürich, Switzerland}

\vspace*{0.5cm}
\begin{center}
\texttt{\small alexander.monin@epfl.ch}  \\
\texttt{\small david.pirtskhalava@epfl.ch}\\
\texttt{\small riccardo.rattazzi@epfl.ch}\\
\texttt{\small fseibold@itp.phys.ethz.ch}
\end{center}

\vspace{.5cm}
\begin{abstract}
\noindent
Hellerman et al. (arXiv:1505.01537)  have shown that in a generic CFT the spectrum of operators
carrying a large U(1) charge can be analyzed semiclassically in an expansion in inverse powers of the charge.
The key is the operator state correspondence by which such operators are associated with a finite density
superfluid phase for the theory quantized on the cylinder. The dynamics is dominated by the corresponding Goldstone
hydrodynamic mode and the derivative expansion coincides with the inverse charge expansion.  We illustrate and further  clarify  this situation by first 
considering simple quantum mechanical analogues. We then systematize the approach by  employing the coset construction
 for non-linearly realized space-time symmetries. Focussing on CFT3 we illustrate the case of  higher rank and non-abelian groups
and  the computation of higher point functions.  Three point function coefficients turn out to satisfy universal scaling laws and correlations  as the charge and spin are varied.

\end{abstract}

\newpage

\setcounter{tocdepth}{2}
\tableofcontents
\newpage
\renewcommand*{\thefootnote}{\arabic{footnote}}
\setcounter{footnote}{0}


\section{Introduction}
Conformal Field Theory (CFT) is essential to describe critical condensed matter systems and  relativistic quantum field theories
in their infrared or ultraviolet asymptotic regimes. Any idea or method offering an insight into the structure of CFTs should therefore be considered of great value. Among the concepts that offer such an insight in specific classes of CFTs we can enlist perturbation theory \cite{Banks:1981nn}, the $\epsilon$-expansion \cite{Wilson:1972cf}, supersymmetry (as elucidated for instance by ref. \cite{Seiberg:1994pq})
and the AdS/CFT correspondence \cite{Maldacena:1997re}. It should also be added that there are many examples of 2D CFTs that are exactly solvable. The conformal bootstrap \cite{Ferrara:1973yt,Polyakov:1974gs}  is instead potentially applicable to CFTs under broader assumptions, though its most spectacular results to date have been obtained in specific systems, such as the 3D Ising model~\cite{El-Showk:2014dwa}. While the majority of the applications of the bootstrap have been relying on numerical methods, some remarkable analytic results have appeared. In particular the bootstrap in the eikonal limit has been used to obtain precise analytic results on the spectrum of operators at large spin $\ell$, showing that in this regime physical quantities follow a semiclassical behavior where $1/\ell$ controls higher order quantum corrections. The possibility to generally describe semiclassically the regime where some charge, not necessarily spin, becomes large has been further elucidated and explored in
ref.~\cite{Hellerman:2015nra}, using a Lagrangian approach. In particular,
focussing on $3$-dimensional CFT, it was
 shown that in the sector of large internal $U(1)$ charge the properties of the  lowest dimension  scalar operators can be studied by considering the system on a spatial 2-sphere in a superfluid state with constant charge  density. We think the general set up and methodology presented in ref.~\cite{Hellerman:2015nra} have a potentially rich range of applications, extending beyond the interesting  but specific results presented in that paper.

The basic picture underlying the analysis of ref.~\cite{Hellerman:2015nra} is the following. By virtue of the operator/state correspondence, a scalar operator with $U(1)$ charge $Q$ corresponds to a state with homogeneous charge density in the theory compactified on the cylinder,  $\mathbb{R}\times S ^ {d-1}$ for a   CFT in $d$ dimensions. Indicating by $R$  
 the radius of the cylinder, the state will have charge density $\rho \sim Q/R^{d-1}$, so that in the limit $Q\gg 1$ there exists a parametric separation between the scale of compactification $1/R$ and the scale associated with the charge density: $\rho^{1/(d-1)} \sim {Q}^{1/(d-1)}/R\gg 1/R$.
 In this window of energy the CFT state and its excitations will therefore correspond to some ``condensed matter phase". As for instance emphasized in ref.~\cite{Nicolis:2015sra}, such phases can, on general grounds, be characterized by the spontaneous breakdown of spacetime and internal symmetries, with their collective excitations dictated by  Goldstone's theorem.
The simplest option, assumed in  ref.~\cite{Hellerman:2015nra}, is that the CFT is in a superfluid phase. This corresponds to a specific pattern of spontaneous symmetry breaking \cite{Nicolis:2015sra} 
 where the $U(1)$ as well as time translations are  broken and 
 just one Goldstone collective excitation boson is mandated. Under these circumstances it is then  possible to systematically compute physical quantities, such as correlators around this state, using the effective Lagrangian for the Goldstone mode(s). The derivative and loop expansion are controlled by powers of the ratio between the IR scale, $1/R$, and the UV scale, $\rho^{1/(d-1)}$,  which correspond to inverse powers of the charge $Q$.
 Order by order in this expansion, the non-universal features associated with any specific CFT will be encapsulated by finitely many coefficients 
 in the effective Lagrangian. The situation is quite analogous to that  of the pion Lagrangian in low energy QCD. In that case  the UV and IR scales are represented respectively by the hadron mass scale $m_{QCD}\sim  4\pi f_\pi \sim 1 $ GeV and by the pion mass $m_\pi \sim 0.1$ GeV and physical observables are controlled by a systematic expansion in powers of $m_\pi/4\pi f_\pi$.

Based on the above picture, in ref.~\cite{Hellerman:2015nra} the spectrum of operators was shown to be calculable, for large  $U(1)$ charge $Q$  and for finite spin $\ell$, as an  expansion in $1/Q$.
 This is undoubtedly already a very interesting result, but there are more directions along which the implications of the method can be generalized and deepened.
One obvious direction to explore is that of large charges for more general groups, including possibly and most interestingly the spin $\ell$. Another direction concerns the computation of correlators.  In the regime of validity of the semiclassical approximation, any operator can be described by expressions with matching quantum numbers  purely written in terms of the Goldstone modes. This is in full analogy with the case of low-energy matrix elements of QCD operators, which are saturated by their expressions in terms of pions.
Particularly interesting is the case of conserved currents, where the matching is more constrained.

The present paper serves, on one hand, the perhaps more modest goal of working out in more detail and from a different perspective the general set up. On the other hand, it 
begins the exploration of  CFTs with multiple large charges, including the non abelian case, as well as outline the computation of correlators.
 In particular we will illustrate how the fixed charge path integral, even for a finite volume system, formally corresponds to the study of configurations with spontaneous symmetry breaking, described by effective Goldstone degrees of freedom. We shall elucidate our discussion  with a semiclassical derivation of the well known spectrum of the rigid rotor in the large $\ell$ limit, which epitomizes the methodology. Furthermore we shall systematize
the derivation of the effective action by employing the general CCWZ~\cite{Coleman:1969sm,Callan:1969sn,Salam:1969rq} methodology for spontaneously broken space-time and internal symmetries. We will thus rederive the results of ref  ref.~\cite{Hellerman:2015nra} and extend them to the, in principle more complicated  case of multiple $U(1)$'s and non abelian groups.

This paper is organized as follows. In Section~\ref{sec_rotor} with the help of a simple quantum mechanical example (rigid rotor) we illustrate how the quasiclassical treatment can be used to describe a system in a sector with large charge (corresponding to angular momentum in this case). In Section~\ref{general_strategy} we present the strategy for studying   general CFT states with large charge using the path integral formulation. Section~\ref{sec_coset} is devoted to describing a tool for building Lagrangians with non-linearly realized symmetries based on the symmetry breaking pattern, the CCWZ or the coset construction. In Section~\ref{sec_superfluid} we demonstrate how the construction works for the  specific example of $U(1)$ symmetry, rederiving the results of ref.~\cite{Hellerman:2015nra}. In Sections~\ref{u1timesu1} and~\ref{sec_SO3} we illustrate the generalization to other internal symmetry groups, dealing in particular with $U(1)\times U(1)$ and $SO(3)$. In Section~\ref{npt} we show how the methodology can be applied to extract other CFT data by computing certain 3- and 4-point functions. In Section~\ref{sec_conclusions} we present our conclusion.

\section{An invitation: the fast spinning (rigid) rotor\label{sec_rotor}}
We here want to illustrate the general connection between large charge, semiclassics and effective Goldstone bosons by focussing on a simple toy example:
a non-relativistic particle in a spherically invariant potential. Indeed, to organize the discussion it is worth to first focus on  the even simpler (and well known) limiting  case
of a particle of mass $M$ constrained to move on a 2-sphere  of radius $a$ whose Lagrangian is
\be
\mc L  = \f {I} {2} \l ( \dot \t ^ 2 + \sin \t ^ 2 \dot \varphi ^ 2 \r ) ,
\label{guess_S2}
\ee
where $I={M a ^ 2}$ is  the moment of inertia. This system is readily exactly solved. The energy eigenfunctions are the spherical harmonics $Y_{\ell m}(\theta,\phi)$ corresponding to energy eigenvalues $E_\ell=\ell(\ell +1)/2 I$. It is interesting to derive this result semiclassically at large angular momentum, by  expanding the path integral around a configuration with $J_3=m\gg 1$.  Notice that in the subspace with fixed $J_3 = m$, the ground state has total angular momentum $\ell = m$ and thus its energy satisfies
 \be 
 E_0(m)=m(m +1)/2 I
 \label{exact}
 \ee
The starting point of our derivation is the standard Euclidean representation of the path integral
\be
\la \t _ f, \vp _ f | e ^ {- H  (\tau _ f - \tau _ i)} | \t _ i, \vp _ i \ra =\int \limits _ {\t,\vp (-\tau _ i)=\t_i,\vp _ i} ^ {\t,\vp (\tau _ f) = \t _ f,\vp _ f}
\mc D \t \mc D \vp e ^ {- \textstyle \int d\tau \, \mc L }
\label{PI}
\ee
with $H $ the Hamiltonian associated to   eq.~(\ref{guess_S2}). Starting from \ref{PI}  we can consider  the matrix element between  eigenstates of the angular momentum $J _ 3 = m$. As $J_3$ and $\vp$ are canonically conjugated  that amounts to
\be
\la \t _ f, m | e ^ {- H (\tau _ f - \tau _ i)} | \t _ i, m \ra = \f {1} {2 \pi} \int \limits _ {\t,\vp (\tau_i)=\t _ i,\vp _ i} ^ {\t,\vp (\tau _ f) = \t _ f,\vp _ f}
\mc D \t \mc D \vp e ^ {- \textstyle \int d \tau \, \mc L } e ^ {-i m (\vp _ f - \vp _ i)} d \vp _ i d \vp _ f .
\label{low_energy_S2_int}
\ee
For $T \equiv \tau _ f - \tau _ i \to \infty$ the amplitude projects onto the lowest energy state $|\Psi_0,m\ra$ with $J_3=m$
\be
\lim_{T\to \infty}\la \t _ f, m | e ^ {- H T} | \t _ i, m \ra=\la \t _ f, m |\Psi_0,m\ra \la\Psi_0,m |\t _ i, m \ra e^{-E_0(m)T}\left [1+ O(e^{-\Delta E(m) T})\right ]
\label{projectm}
\ee
where $\Delta E(m)$ is the energy gap in the $J_3=m$ subspace. The dependence on $T$ and on the coordinates $\theta_{i,f}$ therefore trivially factorizes for $T \D E (m)  \gg 1$.
Now, for $m \gg 1$   the above integral should be computable via a systematic expansion around its saddle points. Including the variation of the boundary terms,  the  stationarity condition is
\be
\ddot \t = \sin \t \cos \t \dot \vp ^ 2, ~~ I \sin ^ 2 \t \dot \vp = -i m~.
\ee
Choosing for instance $\t _ i = \t _ f = \pi /2 $, the solution of minimal Euclidean action  is\footnote{Continuation to real time is done by $\tau = i t$, so that the solution becomes 
$d \vp / d t=m/I$.}
\be
\t _ s = \f {\pi} {2}, ~~ \vp _ s = -i \f {m} {I} \tau +\vp_0\,,
\ee
with $\vp_0$ an integration constant. The solution satisfies 
\be 
\int {\cal L} d\tau +i(\vp_f-\vp_i)m= \frac{m^2}{2I} T
\ee
and is obviously independent of $\vp_0$ as the action only depends on $\dot \vp$. Integration over $\vp_0$ in eq.~(\ref{low_energy_S2_int}) therefore
only trivially affects the overall normalization of the amplitude, and is irrelevant for the computation of the energy eigenvalues. On the other hand,
the integration over the angle $\vp_0$ matters in the computation of correlators involving $\vp$. In particular defining the variables with definite $J_3$ charge
$\psi_q(t)\equiv \exp {iq \vp(\tau)}$ we have 
\be
\la \t _ f, m | \psi_{q_1}(\tau_1)\dots\psi_{q_n}(\tau_n)| \t _ i, m \ra\propto \int_0^{2\pi} \frac{d\vp_0}{2\pi} e^{i\vp_0{\sum_i q_i}}=\delta(\sum_i q_i)
\label{finalint}
\ee
consistent with $J_3$ invariance.

One can also easily check that for $T\to \infty$ the contribution to the stationary action that grows with $T$  is independent on the choice of $\theta_{i,f}$,
as mandated by eq.~(\ref{projectm}). What happens is that for $\theta_{i,f}\not =\pi/2$, along the stationary solution, $\theta(\tau)$ goes exponentially fast towards $\pi/2$ when moving away from $\tau=\tau_i$ and $\tau=\tau_f$.

Rewriting the path integral in terms of the fluctuations $\t = \t _ s+ \x$, $\vp = \vp _ s + \eta$ we have 
\be
Z[T,m]\equiv \la \pi/2, m | e ^ {- H  T} | \pi/2, m \ra =e^{- \f {m^2} {2 I} T} \int\, 
\mc D \x \mc D \eta \, e^{- \int d\tau \l ( \mc L ^ {(2)} + \mc L ^ {\text{(int)}} \r )},
\ee
with
\bea
\mc L  ^ {(2)} & = & \f {I} {2} \l ( \dot \x ^ 2 + \dot \eta ^ 2 + \f {m^2} {I ^ 2} \x ^ 2 \r ), \\
\mc L  ^ {\text{(int)}} & = & \f {m^2} {2 I} \l ( \sin ^ 2 \x - \x ^ 2 \r ) - \l ( \f {I} {2} \dot \eta ^ 2 - i m \dot \eta \r ) \sin ^ 2 \x, \nn 
\label{Ls}
\eea
representing  respectively the free and interacting parts.
By rescaling time $\tau=\tilde\tau /m$, one easily sees that 
\be
\int d \tau \l ( \mc L ^ {(2)} + \mc L ^ {\text{(int)}} \r ) = m\int d \tilde \tau \l ( {{\mc  L '}^ {(2)}} + {\mc L'} ^ {\text{(int)}} \r ),
\ee
where the ${\mc L}'$ correspond to eq.~(\ref{Ls}) with $m=1$. It is thus evident that $1/m$ plays the role of loop counting parameter in 
 the path integral. The computation of all physically relevant quantities will thus be organized as an expansion in inverse powers
of $m$. In particular, for the ground state energy at fixed $J_3=m$ we have
\be
E_0=-\frac{1}{T}\ln Z[T,m]= \sum_{n=0}^{n=\infty} E_0^{(n)}= \frac{m^2}{2I}\l ( 1+ \sum_{n=1}^{n=\infty}a_m m^{-n}\r )
\label{E0series}
\ee
where $a_n$  are $m$-independent coefficients and where we factored out the classical contribution. 

To perform our computation we should also regulate our path integral. Moreover we should better do so compatibly with
the defining symmetries. Our system  can be viewed as the $SO(3)/SO(2)$ $\sigma$-model over a 1D space-time. Using the standard  CCWZ construction to classify the invariants it is then simple to power count the possible divergences arising from the original Lagrangian \ref{guess_S2} compatibly with symmetry\footnote{For readers unfamiliar with the construction and the power counting, we shall recall how this works in a later section.}. 
Not surprisingly, given what we know from the exact solution, the only possible divergence is a linearly divergent
contribution to the cosmological constant, which arises at 1-loop. This corresponds  to a trivial $(\ell,m)$ independent shift of all energy levels. Notice in particular that no renormalization of the inertia moment $I$ is needed. Moreover in dimensional regularization, which is the regulator of choice to respect the $\sigma$-model symmetry, there are no power divergences, so that the path integral is automatically finite.

\begin{figure}[h]
\begin{center}
\includegraphics[height=3cm]{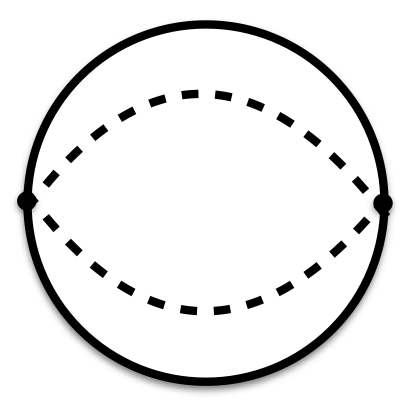} 
\caption{\label{fig_3_loop} Three-loop diagram}
\end{center}
\end{figure}

Let us then consider the lowest order contributions to $E_0$. At 1-loop we have the contribution from the fluctuation determinant associated with the kinetic terms of $\xi$ and $\eta$ in $\mc L  ^ {(2)}$
\be
\Delta ^{(1)}E_0=\frac{1}{2}\int \frac{d\omega}{2\pi} \l (\ln (\omega^2+m^2/I^2)+\ln \omega^2\r)= \Lambda_{UV} +\frac{m}{2I}\,.
\ee
The divergence term of course depends on the regularization, which we have not specified, and in particular vanishes in DimReg.
The finite part corresponds to the zero point energy of the harmonic oscillator $\xi$. Notice that this contribution together with the classical
one $=m^2/2I$  nicely saturates the exact result \ref{exact}. What about higher loops then? 
\begin{figure}[h]
\begin{center}
\includegraphics[height=2cm]{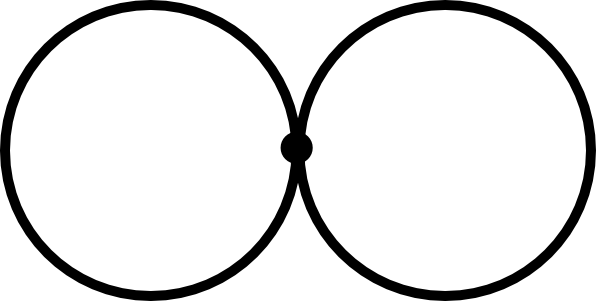} \hspace{1cm}
\includegraphics[height=2cm]{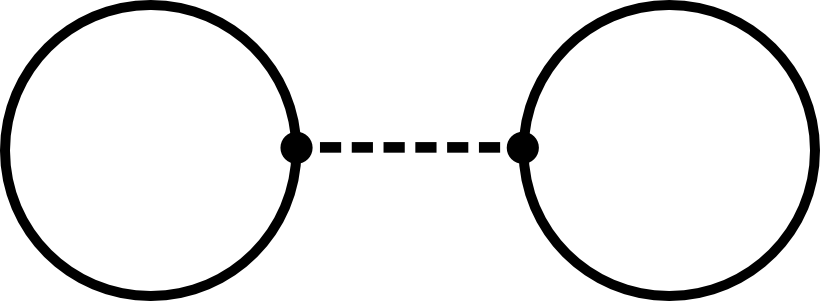} \hspace{1cm}
\includegraphics[height=2cm]{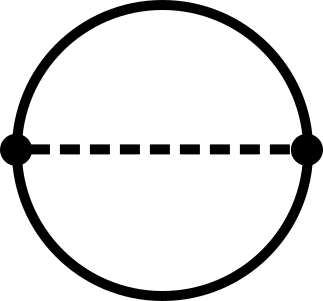}
\caption{\label{fig_two_loop} Two-loop diagrams. Propagators of $\t$ and $\vp$ are represented by solid and dashed lines correspondingly.}
\end{center}
\end{figure}
At two loops  we have the  three diagrams depicted in Fig.\ref{fig_two_loop} (other diagrams are trivially zero in DimReg). Using the expression for the propagators
\be
\l \la \x (\omega) \x (-\omega) \r \ra = \f {1} {I\omega^2+ m^2/I}\ \, ~~ \l \la \eta (\omega)  \eta (-\omega) \r \ra = \f {1} {I \omega^2} 
\ee
and for the vertices these diagrams are all seen to be  manifestly finite and to sum up to 
\be
\D^ {(2)} E _ 0 =   \f {1} {4 I}.
\ee
Combining the contributions up to two loops we then have
\be
E ^{\text{2-loop}}_ 0 = \f {m (m + 1)} {2 I} + \Lambda_{UV} + \f {1} {4 I}.
\label{S2_2loop_saddle}
\ee
As we already mentioned the  $m$-independent constant term cannot be predicted, for there is a UV divergence which scales  precisely as $m ^ 0$. The finite 2-loop contribution is thus inessential. Starting with the next order, 3-loops, the contributions will be suppressed by a positive power of $m$. However, and this 
seems quite remarkable from our perturbative  perspective,  in view of the exact result \ref{exact}, each and every higher order term should exactly vanish! This is far from evident by just looking at the explicit form of the Lagrangian, but it must be so given the underlying $SO(3)$
symmetry. To be reassured that we are not missing anything we have indeed performed the leading non trivial check by computing the 3-loop contribution. Here, unlike at two loops, there are formally divergent diagrams, such as the one in Fig.~\ref{fig_3_loop}, which is naively proportional to $\delta(0)$. It is then crucial to perform a symmetric regulation of the integrals, and the simplest option is DimReg. 
\noindent
In so doing we  checked that at three loops all the diagrams indeed non-trivially sum up to zero. Indeed it is perhaps worth pointing out that  the result is guaranteed by the occurrence  of a crucial evanescent contribution. When extending our system to $d=1+\epsilon$ dimension the term $(\dot \eta)^2$, which is a perfect square in 1D, is extended to $\partial_\mu\eta\partial ^\mu \eta$, which is no-longer a perfect square. Thanks to that, the diagram in Fig.\ref{fig_3_loop}, which is naively proportional to $\delta(0)$ prior to regulation,
is not extended to $\delta^ {(d)}(0)=0$ in DimReg. It instead gives a finite result which crucially ensures the vanishing of the whole 3-loop $O(1/m)$ correction.

The semiclassical description should not be limited to the ground state at fixed charge. It is straightforward to check that is the case at the lowest relevant order, at which the excited states are described by the levels of the $\xi$ harmonic oscillator whose frequency is $m/I$. The state with $n$ quanta thus has a gap  $\Delta_n E= n m/I$ above $E_0$, so that we can write
\be
E_n = E_0+\Delta E_n+O(m^0)= \frac{(m+n )(m+n+1)}{2I}+O(m^0)
\ee
corresponding to the state with $J_3=m$ and $\ell = m+n$. For the excited states, two loop effects should crucially intervene
to match the $m^0$ terms, but we have not checked that.

After having understood the rigid rotor limit, it is worth going back to the general case of a 3D particle in a potential $V(r)$. 
Working in polar coordinates, the Euclidean action describing the path integral at fixed value of $J_3$ is now
\bea
\mc L &=&\frac{M}{2} \l ({\dot r}^2+r^2{\dot \theta}^2+ r^2 \sin \theta^2{\dot\vp}^2\r ) +V(r) +im \dot \vp\, \\
&=& \frac{M}{2} \l [{\dot r}^2+r^2 {\dot\theta}^2+r^2\sin \theta^2\l ({\dot \vp}+i m/(Mr^2\sin^2\theta)\r)^2 \r ] +\frac{m^2}{2Mr^2\sin^2\theta}+V(r)
 \eea
Working at large $m$ we can proceed semiclassically and  expand around the  (leading)  stationary point of the above action. Assuming the effective potential $V_{eff}(r,\theta,m)$ (given by the last two terms in the second equation above) is stationary at $r=r(m)$, for $\theta=\pi/2$,  the solution generalizing our previous one is
\be
\t _ s = \f {\pi} {2}, ~~ \vp _ s =- i \f {m} {Mr(m)^2} \tau +\vp_0\,,~~ r=r(m) \,.
\label{solution2}
\ee
Considering the small fluctuations around this solution, we have that $\vp$ is a zero mode while  the modes $\theta$ and $r$ are massive
with frequencies given respectively by ($I(m)=Mr(m)^2$)
\be
\omega_\theta^2=\frac{m^2}{I(m)^2}\qquad \omega_r^2=\frac{3m^2}{I(m)^2}+\frac{V''(r(m))}{M}\, .
\label{thetaandr}
\ee
For large $m$, and barring cancellation form $V''$, the fluctuations of both $\theta$ and $r$ are therefore small. Qualitatively (using for instance the 
expression for the coordinate fluctuations on the harmonic oscillator ground state) we have
\be
(\Delta \theta)^2 \sim   \frac{1}{m}\qquad \frac{(\Delta r)^2}{r(m)^2}\sim \frac {1}{m}\frac{1}{\sqrt {3+ r(m)^4MV''(r(m))/m^2}}
\label{massr}
\ee
Assuming $V$ to be sufficiently generic, we have that the larger $m$, the more $r$ is localized away from zero over the dominant trajectories in the path integral. This is intuitively expected: at fixed large $m$ the centrifugal force keeps $r$ away from the point $r=0$ where spherical symmetry is classically restored. The large value of $m$ forces the path integral to be dominated by small fluctuations around the classical solution \ref{solution2}, which like
most classical solutions ``spontaneously breaks" the exact symmetries of the problem. In the present case the symmetry is given by rotations  
and time translation $SO(3)\times \cal T$, and the pattern of breaking induced by \ref{solution2} is $SO(3)\times {\cal T}\to {\cal T}'$, where the  generator  of the unbroken time translation ${\cal T}'$ is given  (in an obvious notation) by ${H}'= H-(m/I) J_3$ \footnote{Things are more transparent in real time $t\to it$. In euclidean time one must onsider an unusual, but perfectly fine, imaginary time translation.}. Moreover, the Goldstone velocity $d \vp / d t =m/I$, which is canonically related to $J_3$, plays expectedly  the role of  chemical potential $\mu$.
This pattern of symmetry breaking, where time translations mix with an internal symmetry, is the simplest option to give rise to a configuration with  finite charge density. We should however stress  here that, while the generators $J_1$ and $J_2$ are plainly broken by our choice of boundary conditions with fixed $J_3$, the spontaneous breaking of $J_3$ and $\mc T$ is instead just a property of the solutions, parametrized by $\vp_0$, that dominate the path integral. As already indicated by the discussion around eq.~(\ref{finalint}), integration over $\vp_0$ properly enforces in the end the invariance of the correlators under the action of $J_3$ and time translations.
This of course has to be the case as in a  quantum mechanical system such as the one at hand there cannot be spontaneous symmetry breaking.
Yet, modulo the final integration on the zero mode $\vp_0$, in all the stages of the computation it is technically correct and useful to view the symmetry as spontaneously broken. Precisely the same remarks apply to the CFT compactified on the sphere we shall consider later: while at finite volume 
there is strictly speaking no spontaneous symmetry breaking, boundary conditions and the semiclassical method will effectively enforce it.

One final issue, which will be  useful to draw analogies in the CFT case, concerns the gap of the radial mode and the possibility to integrate it out.
Eq.~(\ref{thetaandr}) tells that $\theta$ and $r$ have comparable frequencies of the order of the chemical potential\footnote{Indeed the frequency of the Goldstone $\theta$ is precisely fixed to be equal to $\mu$ by the $SO(3)$ algebra (see for instance \cite{Nicolis:2012vf}).} $\mu \sim m/I$, unless $V$ contains some specific large parameter.
To get a quick idea it suffices to consider a powerlike potential $V(r)=cr^\alpha$ in which case $\omega_r^2=(\alpha+2) m^2/I^2$. The parameter $\alpha$ 
thus controls the rigidity of the rotor: for  $\alpha \gg 1$ we  have $\omega_r\gg \omega_\theta$ and $r$ can be integrated out to effectively describe the system as a rigid rotor term plus a  series of $1/\alpha$ suppressed higher derivative terms describing deviations from perfect rigidity. In the case $\alpha \gg 1$ we can thus describe the system in terms of the pure $SO(3)/SO(2)$ $\sigma$-model, but generically we expect degrees of freedom with comparable mass to those of $\theta$ dictated by the $\sigma$ model over the rotating solution. This situation occurs  also in the case of genuine field theories that we shall consider later. However in that case the dynamics of the massless modes, the analogues of $\eta$, is more consequential than in the case of the rigid rotor. The truly robust predictions in the CFT case concern the latter degrees of freedom, as it will become clear later on.

\section{Path integral at fixed charge and Goldstone bosons\label{general_strategy}}
The  approach of ref.~\cite{Hellerman:2015nra} can be viewed as a field theoretic  version of the quantum mechanical example of the previous section. 
Our  goal is to present the results of ref.~\cite{Hellerman:2015nra} from a different perspective and to extend them to the case of multiple, possibly non abelian, charges. 
Considering  a general $d$-dimensional CFT with a global (internal) symmetry group ${\cal G}$ of rank $N$ we want to study the properties of primary operators  ${\cal O}_{\vec Q, a}$
carrying large values of the conserved charges $\vec Q=(Q_1,\dots, Q_N)$ associated to the Cartan generators $\widehat Q_I$. Here the index $a$ labels dimension,  spin  and possibly extra discrete quantum numbers.
In particular, working on the Euclidean plane $\mathbb{R}^d$, the goal is to systematically distill the universal properties of correlators of the form
\be
\la {\cal O}_{-\vec Q,a}(x_{out}){\cal O}_m(x_m)\dots{\cal O}_1(x_1){\cal O}_{\vec Q,a}(x_{in}) \ra
\label{planecorr}
\ee
where by ${\cal O}_{-\vec Q,a}$ we indicate the operator corresponding to the Hermitian conjugate of ${\cal O}_{\vec Q, a}$ in the Minkowskian continuation\footnote{In the Euclidean theory  with radial quantization, considering 
for instance a scalar primary of dimension $\Delta$,  we have instead the relation ${\cal O}_{\{Q\}}(x)^\dagger=x^{-2\Delta} {\cal O}_{\{-Q\}}(\hat Rx)$ where $\hat Rx$ is the image of $x$ under space inversion (see for instance ref.~\cite{Pappadopulo:2012jk}).} 
while the ${\cal O}_j$'s are operators with finite  values of all the quantum numbers, including the $Q_I$. For instance they could include the energy momentum tensor and the conserved ${\cal G}$ currents. In order to proceed it is convenient to map to the cylinder $\mathbb{R}\times S ^ {d-1}$ and exploit the operator state correspondence. In polar coordinates $x\equiv (r=|x|,{\bf n})$ is mapped to $(\tau=R \ln r / R, {\bf n})$, where $R$ is the radius of the $S ^ {d-1}$ sphere. Normally units where $R = 1$ are chosen, but for later purposes (dimensional analysis) we keep the radius arbitrary. 

Modulo Jacobian factors that are fully determined by the dimension and spin of the ${\cal O}$'s\footnote{For instance in the case of a scalar primary of dimension $\Delta$ one has ${\cal O}(\tau,{\bf n})_{\mathrm {cyl}}=(r/R)^\Delta 
{\cal O}(r,{\bf n})_{\mathrm {{\mathbb{R}}^D}}$. When mapping to the cylinder there is in general a Weyl anomaly. However the effects of the latter are a global identical shift of all energy levels on the cylinder  together with ultralocal contributions to the correlators (in particular for the energy momentum tensor). It therefore does not affect the discussion as long as we only consider correlators at non-coincident points normalized by the vacuum to vacuum amplitude.},  eq.~(\ref{planecorr}) is given by the corresponding correlator on the cylinder. For  $x_{in}\to 0$ and $x_{out} \to \infty$ (i.e. $\tau_{in}\to -\infty$ and $\tau_{out}\to \infty$),  the action of ${\cal O}_{\vec Q, a}$  projects on the lowest energy eigenstate $|\vec Q,a\ra$ 
in the subspace spanned by ${\cal O}_{\vec Q,a}(x)|0\ra$
\bea
\lim_{\tau_{in}\to -\infty} {\cal O}_{\vec Q, a} (\tau_{in},{\bf n}_{in})\l.| 0\r\ra&=&e^{E_{\vec Q, a} \tau_{in}}|{\vec Q},a\ra \equiv |\vec Q,a,\tau_{in}\ra\\
\lim_{\tau_{out}\to \infty} \la0|{\cal O}_{-\vec Q, a} (\tau_{out},{\bf n}_{out})&=&\la \vec Q, a|  e^{-E_{\vec Q, a} \tau_{out}}\equiv \la \vec Q, a, \tau_{out}| \, ,
\eea
with $E_{\vec Q, a} = \Delta_{\vec Q, a} /R$ and $\Delta_{\vec Q, a}$ the corresponding dimension of $\mc O _ {\vec Q, a}$. The computation of eq.~(\ref{planecorr}) is then equivalent to the computation of
\be
\la \vec Q, a, \tau_{out}| {\cal O}_m(\tau_m, {\bf n}_m)\dots{\cal O}_1(\tau_1,{\bf n}_1)|\vec Q, a,\tau_{in}\ra\,
\label{expect}
\ee
on the cylinder. We are here is a situation quite analogous to that of the previous section. It is thus reasonable to assume that, at large $Q_I$'s, the path integral
computation of the above quantity will be dominated by semiclassical trajectories characterized by a specific pattern of symmetry breaking.
The trajectory with lowest action will be associated with the state $|{\vec Q}\ra$ of lowest energy $\Delta_{\vec Q}$ in the subspace with fixed $\vec Q$. We shall indicate  by ${\cal O}_{\vec Q}$ the  specific operator corresponding to such ``ground state".
Operators/states with higher energy will correspond to the excitations around the lowest action  trajectory.
Such a leading trajectory must have   the same symmetry properties associated to the two  insertions of ${\cal O}_{\vec Q}$ at respectively $x_{in}=0$ and $x_{out} = \infty$. This is because these insertions set the boundary conditions for the path integral.
 As concerns the conformal group, the insertion at $0$ breaks
translations $P_\mu$, while the insertion at  $\infty$ breaks  special conformal $K_\mu$.  Rotations on the sphere $SO(d)$ may or may not be broken
depending on whether ${\cal O}_{\vec Q}$ has a spin. In what follows we shall assume  ${\cal O}_{\vec Q}$ is a scalar, corresponding to the rather plausible situation where the ground state and the leading trajectory are rotationally invariant. As argued in ref~\cite{Hellerman:2015nra}, and as we shall repeat later, one can actually quantitatively prove the self-consistency of this assumption. There just remains one generator of the conformal group whose fate we must debate : $D$, generating   dilations on the plane and time translation on the cylinder. Now, the points $x_{in}=0$ and $x_{out} = \infty$ are stable under dilations, corresponding to $|\vec Q\ra$ being an eigenstate of time evolution on the cylinder. We thus expect that the leading trajectory should therefore be invariant under an effective time translation operator $D'$. 
 On the other hand, as $|\vec Q\ra$ carries $\cal G$ charges, the leading trajectory  will generically  only respects a subgroup ${\cal H}\subset {\cal G}$, and possibly, like in the case of the rigid rotor, ${\cal G}$ will be fully broken. We conclude that the trajectory will be characterized by a symmetry breaking pattern $SO(d+1,1)\times {\cal G}\to SO(d)\times D'\times {\cal H}$, in an obvious notation.  In view of that  the fluctuations around the background will necessarily count
a set of Goldstone bosons whose effective action is largely dictated by symmetry considerations. However, like for the rotor's radial mode,
there could possibly exist additional light degrees of freedom. While this option in some specific cases may be dictated by additional symmetries, such as supersymmetry, we shall first work  by assuming  there exists a gap between the Goldstones and the other excitations. The latter can then be meaningfully  integrated out describing the system via a  general effective action for the Goldstone bosons. We shall later come back and consider in more detail the assumption of a large gap: as it turns out, in the case of a non-abelian ${\cal G}$, it needs to be better qualified.

The leading semiclassical solutions will thus correspond to a  homogeneous state on $S^{d-1}$, characterized by  large charge   densities $\rho_I= Q_I/R ^ {d-1}\mathrm{Vol}_{S^{d-1}}$, where $\mathrm{Vol}_{S^{d-1}}$ is the volume of a unit $S ^ {d-1}$ sphere. The simplest option for the pattern of symmetry breaking corresponding to such a state is given by a ``generalized superfluid". That is defined as the situation where time translations $D$ as well as at least one linear combination of charges $\hat \mu_I \widehat Q_I$ are spontaneously broken, but where there remains an effective unbroken ``diagonal" time translation $D'=D+\mu_I \widehat Q_I$. Again, this is precisely the situation we encountered in the case of the rotor
(see discussion below eq.~(\ref{massr})).
We expect that for generic choices of the $Q_I$ the  pattern of breaking that realizes this state of things features no residual unbroken symmetry ${\cal H}$.  However, for specific directions in $Q$ space, there exists the option to have a residual symmetry~\cite{Alvarez-Gaume:2016vff}. Consider for instance ${\cal G}=SO(2n)$, where we can conveniently associate the $Q_I$ to the block diagonal generators $\widehat Q_1=(\epsilon, 1,\dots, 1)$, $\widehat Q_2=(1,\epsilon, 1,\dots,1)$, etc. Then the case $Q_1\gg 1$ and $Q_{I\geq 2}=0$ will clearly be compatible with a background respecting a residual $SO(2n-2)$. Indeed, and as already pointed out in ref.~\cite{Hellerman:2015nra}, one could consider different realizations of a homogeneous state with large charge density, 
as for instance offered by a fermi liquid. As spacetime symmetries in a fermi liquid are broken, some of the bosonic excitations must have the interpretation of Goldstone bosons, but certainly the construction does not follow the same universal rules of purely bosonic systems, so it is less clear to us how to proceed in general.  Notice indeed that  this situation is not captured by the coset classification of ref.~\cite{Nicolis:2015sra}. For that reason we will focus on systems, such as purely bosonic ones, where the leading solution is a generalized superfluid.

One last comment concerns a more direct interpretation of the pattern of symmetry breaking. That is gained by taking the formal limit 
where the radius ${R}$ is sent to infinity to recover ${\mathbb{R}}^d$. We denote the conformal group generators acting in $\mathbb R ^ d$ obtained by that procedure  by 
\be 
\wh P _ \m,~~\wh J _ {\m \n}, ~~\wh D, ~~ \wh K _ \m\, .
\label{local_conformal}
\ee
As shown in Appendix~\ref{R_infty_limit} they can be straightforwardly mapped into the original ones.  The broken original generators $P_\m$, $K _ \m$ and $D$
are mapped into certain combinations of $\wh J _ {0i}$, $\wh D$, $\wh K _ \m$ and $\wh P _ \m$. In particular, ordinary dilations are generated by $\wh D = R P_0 /2 - K_0 / 2 R$ in this limit. On the other hand, the unbroken generators $SO(d)\times D'$ are mapped to the $d$-dimensional Euclidean group (spatial rotations $\wh J_{ij}$ and translations $\wh P_j$) plus effective time translations generated by 
$\wh H' \equiv \wh P' _ 0\equiv \wh P_0+\mu_I \wh Q_I$. This is the symmetry of homogeneous and isotropic condensed matter \cite{Nicolis:2015sra}. Conformal invariance and boosts are spontaneously broken as there exists a finite charge density $\rho_I$ (and the corresponding finite energy density).

In the next section we will recall the general methodology to write down effective Goldstone Lagrangians with non-linearly realized space-time symmetry
and adapt it to the case of a {\it {generalized conformal superfluid}} $SO(d+1,1)\times {\cal G}\to SO(d)\times D'\times {\cal H}$. The path integral 
will be written as a generalization of the rigid rotor example. Consider for  simplicity the case where ${\cal G}$ is fully broken. Among the set of Goldstones $\{ \chi\}$, there will be the subset of  $N$ Goldstones $\chi^I$ associated with the Cartan generators.
The path integral around the vacuum state $|\vec Q\ra$ will therefore be written as
\be
\la \vec Q| e ^ {- \wh H T} | \vec Q\ra = \int d ^ N \c _ i \, d ^ N \c _ f \,  \int _ {\ba{l}  \scriptstyle\c (\tau _ f) = \c _ f \\ \scriptstyle \c (\tau _ i) = \c _ i \ea} \mc D \c \exp \l(  {-S\l [\c \r] - \f {i} {\mathrm{Vol}_{S^{d-1}}}\int d\tau d^{d-1}{\bf n} \,Q _ I \dot \c ^ I } \r ).
\label{GoldstoneAmplitude}
\ee
where as before $T = \tau _ f - \tau _ i$ and we have assumed a parametrization where each  Goldstone $\c ^ I$ is canonically conjugated to the charge $Q_I$. In view of that, and in full analogy with the rotor example,  the last term in the action  acts  as a wave functional projector on initial and final states with suitably fixed charges.
In the limit $Q _ I \gg 1$ the above integral  can be computed via the saddle point method with $1/Q_I$ controlling the loop expansion. 

Eq.~(\ref{GoldstoneAmplitude})  can be used to derive a relation between $\vec Q$ and the energy $\Delta_{\vec Q}/R$ of the ground state. However, it can be obviously generalized to compute other quantities. In particular, in the regime of validity of the above effective action, which we shall elaborate upon later, any other operator of the CFT can be  represented in terms of the Goldstone bosons $\chi$ just by matching its $SO(d+1,1)\times {\cal G}$ quantum numbers. In particular, the energy momentum tensor will be matched by the energy momentum tensor of the Goldstone action.

The crucial step is the construction of the most general effective action $S[\chi]$, consistent with the desired symmetry breaking pattern -- something we turn to discussing next.


\section{The coset construction \label{sec_coset}}

In this section we illustrate the general methodology for constructing invariant Lagrangians for the explicit case of ${\cal G}=U(1)$. The latter example has been already discussed in 
ref.~\cite{Hellerman:2015nra}, but it will serve us the purpose to introduce the general Callan-Coleman-Wess-Zumino (CCWZ) construction~\cite{Coleman:1969sm,Callan:1969sn,Salam:1969rq} for non-linearly
realized spacetime symmetries, which almost straightforwardly generalizes to arbitrary groups ${\cal G}$. It should be perhaps noted, that the CCWZ construction is not so much needed to construct the leading order Lagrangian in the simplest case at hand, but it is crucial to properly control the systematic expansion once higher-order effects are probed.

\subsection{Non-linearly realized internal symmetries\label{CCWZ}}
By Goldstone's theorem, the spontaneous breakdown of a global symmetry ${\cal G}\to {\cal H}$ implies that the low-energy physics is described by Goldstone bosons spanning
the coset space  ${\cal G}/{\cal H}$. The CCWZ method allows to construct the most general interaction Lagrangian for these modes.  Indicating by $X_\a$ the unbroken generators and by $T_a$ the broken ones, the coset space is parametrized by 
\be
\Omega = e ^ {i \pi^a T_a} \in {\cal G} \ ,
\label{coset_rep}
\ee
where $\pi^a$ are the Goldstone fields in one-to-one correspondence with the broken generators.
The transformation of the  $\pi ^ a$'s under the action of an element of the global group $g \in {\cal G}$ is  given by
\be
g\Omega = \Omega ' h\equiv e^{i\pi' T} h\ , ~~~\text{with}~~~  h\equiv h (\pi,g) \in {\cal H} \ .
\label{g_action_coset}
\ee
As a next step, one considers the \textit{Maurer-Cartan 1-form}
\be
\Omega ^ {-1} \p _ \m \Omega = i \nabla _ \m \pi ^ a \, T _ a + i \Xi _ \m ^\a X_\a \ ,
\label{MC_form}
\ee
where the dependence of the coefficients $\nabla _ \m \pi ^ a$ and $\Xi _ \m ^ \a$ on the $\pi^a$ is fixed by the algebra of the group. In particular, one has $\nabla _ \m \pi^a=\partial_\m\pi^a+\dots$~. The Maurer-Cartan form transforms in the following way
\be
\l ( \Omega ^ {-1} \p _ \m \Omega\r ) \to \l ( \Omega ^ {-1} \p _ \m \Omega\r ) ' = h \l( \Omega ^ {-1} \p _ \m \Omega \r ) h ^ {-1}  + h \p _ \m  h ^ {-1} ~,
\ee
which  can be equivalently rewritten as
\be
\begin{aligned}
\nabla_\m \pi ' T & =  h  \, \nabla_\m \pi ' T \, h^{-1}  \ ,  \\
i\Xi _ \m ' X & = h \, i\Xi_ \m  X \, h^ {-1} + h\p _ \m h^ {-1} \ .
\label{h_trans_coset}
\end{aligned}
\ee
The crucial remark is that the action of global $g\in {\cal G}$ is described on the Goldstones by a local $ h(\pi,g)\in {\cal H}$. The \textit{covariant derivative}~$\nabla_\m \pi$ transforms linearly under such a local ${\cal H}$, while $\Xi _ \m$ transforms like an ${\cal H}$-gauge field. For this reason, $\Xi _ \m$  can be used for coupling Goldstones to other fields or for constructing higher-derivative covariant operators. Indeed, given a field $\psi$ living in a $k$-dimensional representation $\rho$ of $\mc H$
\be
\mc H \ni h: ~ \psi ' = \rho (h) \psi,
\ee
it is easy to check that the derivative $D _ \m \psi$ defined according to
\be
D _ \m \psi = \p _ \m \psi + \Xi _ \m ^ \a \rho (X _ \a) \psi,
\ee
transforms covariantly,
\be
(D _ \m \psi) ' = \rho (h) \p _ \m \psi  + (\p _ \m \rho (h) ) \psi + \Xi _ \m ^ { '\a} \rho (X _ \a) \psi  = \rho (h) D _ \m \psi.
\ee
As a result, any invariant under the local ${\cal H}$ is automatically also invariant under the action of the global ${\cal G}$. This makes the construction of invariant Lagrangians a rather straightforward task.

One important aspect of the CCWZ construction is the possibility to lift  $\mc H$ representations to corresponding $\mc G$ representations.
Consider indeed a field $\psi$ in a $k$-dimensional representation $\rho$ of  $\mc H$ which appears in the decomposition of a $K$-dimensional representation $\kappa$ of $\mc G$.
Defining 
\be
\label{reps1}
 \tilde \psi = \underbrace{( \psi, \overbrace {0} ^ {K - k} )} _ {K},
\ee
it is easy to show that the field $\Psi \equiv \kappa (\Omega) \tilde \psi$ transforms linearly under the group $\mc G$
\be
\label{reps2}
\Psi ' = \kappa (\Omega ') \tilde \psi ' = \kappa (g \Omega h ^ {-1}) \kappa (h) \tilde \psi = \kappa (g) \kappa (\Omega) \tilde \psi = \kappa (g) \Psi.
\ee

The CCWZ construction generalizes straightforwardly to the case when some of the symmetries are gauged. This is achieved by promoting the partial derivative in the Maurer-Cartan form to a covariant one through the inclusion of gauge fields that transform under ${\cal G}$ in the standard way, $
\tilde A ' _ \m = g \tilde A _ \m g ^ {-1} + g \p _ \m g ^ {-1} \ .$

\subsection{Non-linearly realized space-time symmetries}

The CCWZ construction of the previous subsection straightforwardly generalizes to a situation where one or more \textit{spacetime} symmetries are broken on top of the internal ones. We will avoid giving a formal discussion of how the coset construction works for the most general case with nonlinearly realized spacetime symmetries. Instead, we will illustrate the construction on a series of increasingly involved examples, eventually arriving at the one relevant for the generalized superfluid we are interested in in this paper. We will discuss the latter case quantitatively, while giving a more sketchy description of the preceding examples (intended mainly for providing an invitation to the subject).

\subsubsection*{General Relativity}

A prototype example of the CCWZ construction for nonlinearly realized spacetime symmetries is General Relativity, as  viewed from the coset perspective~\cite{Ivanov:1981wn,Delacretaz:2014oxa}. The relevant coset in that case is $ISO(3,1)/SO(3,1)$, corresponding to the tangent-space Poincar\'e group with non-linearly realized translations and linearly realized (local) Lorentz transformations. The coset is thus parametrized by 
\be
\Omega = e^{iy^a(x) \wh P_a}~,
\ee
where $y^a(x)$ are the tangent space coordinates that in the given approach play the role of the Goldstones, corresponding to `spontaneously broken' translations (we will denote all generators acting in a local chart of the base manifold by hatted symbols).\footnote{Since $y^a$ are associated with the spontaneously broken \textit{local} generators, they are analogous to the St\"uckelberg fields that can be chosen at will due to the gauge redundancy.} From now on, the indices $a,b, \dots=0,\dots, d-1$ will label the gauged Poincar\'e generators, and should be distinguished from the space-time indices. Notice that the action of diffeomorphisms accounts to a mere relabeling of the space-time coordinates $x^\mu$ (which do not transform under the tangent-space translations $\wh P_a$). 

Just as for spontaneously broken internal symmetries, the next step is to define the Maurer-Cartan one-form, introducing the gauge fields $\tilde e^a_\mu$ and $\omega^{ab}_\mu$ for the local translations and Lorentz transformations
\be
\Omega^{-1}D_{\mu}\Omega =  e^{-iy^a(x) \wh P_a} \(\p_\mu+i\tilde e^a_\mu \wh P_a+\frac{i}{2}\omega^{ab}_\mu \wh J_{ab}\)   e^{iy^a(x) \wh P_a} = i e^a_\mu \wh P_a+\frac{i}{2}\omega^{ab}_\mu \wh J_{ab}~.
\ee
In the last step we have defined $e^a_\mu = \tilde e^a_\mu+\p_\mu y^a+\omega^{ab}_\mu y_b$, which, according to the discussion around eq.~\eqref{h_trans_coset}, transforms covariantly under all symmetries.  For that reason, $e^a_{\mu}$ is naturally identified with the usual vielbein of General Relativity, and we will see shortly that it is indeed the standard `square root' of the metric, $g_{\mu\nu}=\eta_{ab} e^a_\mu e^b_\nu$; in particular, it can be used to construct an invariant volume element $d^d x \det e$. In contrast, $\omega^{ab}_\mu$ transforms like a $SO(3,1)$ gauge field (the spin connection) and can be used to couple matter fields to gravity, as well as construct higher-derivative covariant operators. 

Note, that at this stage both $e$ and $\omega$ are independent fields. This is in contrast to GR, where the spin connection is (algebraically) expressed in terms of the vielbein and its derivatives. To see how the latter relation arises from the CCWZ perspective, consider the \textit{curvature two-form}
\be
\label{curvature2form}
\Omega^{-1} [D_\mu,D_\nu]\Omega \equiv i T^a_{\mu\nu} \wh P_a + \frac{i}{2} R^{ab}_{\mu\nu} \wh J_{ab}~,
\ee
where $T^a_{\mu\nu}$ and $R^{ab}_{\mu\nu}$ can be schematically written as $T\sim \p e+\omega e$ and $R\sim \p\omega +\omega\omega$. By construction, both of these objects transform linearly under all symmetries, which allows to define a reduced theory that satisfies a covariant constraint $T^a_{\mu\nu}=0$~. This is the usual torsion-freedom condition of General Relativity that removes the spin connection as a dynamical degree of freedom. We will see more examples of such constraints, which are often referred to as the \textit{inverse Higgs constraints} in the CCWZ literature (see Sec.~\ref{constrs} for more on this). After expressing the spin connection in terms of the vielbein and its derivatives, $R^{ab}_{\mu\nu} e_a^\mu e_b^\nu$ reduces to the standard Ricci scalar which one can use to build the (Einstein-Hilbert) action for the dynamical graviton.

Any extra field with definite transformation properties under the tangent-space symmetries can be straightforwardly coupled to the gravitational degrees of freedom in a way outlined in sec.~\ref{CCWZ}. A theory without gravity on a fixed (in general curved) background manifold is obtained by freezing the vielbein to the value $\la e^a_{\mu}\ra$, corresponding to a particular choice of the coordinate system on that manifold. The isometries, if any, of the resultant theory are then characterized by diffeomorphisms that, when acting on $\la e^a_{\mu}\ra$, reduce to a Lorenz transformation that can be undone by an element of the Local Lorentz group (in other words, the isometries belong to the subgroup of local Lorentz $\times$ diffs that leaves $\la e^a_{\mu}\ra$ invariant).

\subsubsection*{Dilaton gravity}

Our next example corresponds to the case where one endows the above theory with a non-linearly realized local dilation invariance on top of the local Poincar\'e group. Not surprisingly, going through similar steps as we did for GR  yields in this case a Weyl-invariant theory of the dilaton, coupled to gravity. To see this, consider the following parametrization of the coset
\be
\label{omdilgrav}
\Omega = e^{iy^a(x) \wh P_a} e^{i\sigma \wh D}~,
\ee 
where $\sigma$ is the Goldstone boson, non-linearly realizing the dilations (the dilaton).
 Introducing the gauge field for dilations, $A_\mu$, eq.~\eqref{omdilgrav} yields the following expression for the Maurer-Cartan form (see~\cite{Karananas:2015eha, Karananas:2016hrm} for details)
\bea
\Omega ^ {-1}  D _ \m \Omega & = & \Omega ^ {-1} \l( \p _ \m + i \tilde e _ \m ^ a \wh P _ a + \f {i} {2} \omega ^ {ab} _ \m \wh J _ {ab} +i  A_\m \wh D \r)\Omega  \\
& = & e^{-i\sigma \wh D} \l( \p _ \m + i e _ \m ^ a \wh P _ a + \f {i} {2} \omega ^ {ab} _ \m \wh J _ {ab} +i  A_\m \wh D \r)  e^{i\sigma \wh D} \\
& = & i E_\m^a \l(\wh P_a + \nabla_a \s \, \wh D + \f{1}{2}e^\sigma \om_a^{bc}\wh J_{bc} \r) ,
\eea
where 
\be
E^a_\mu = e^{-\sigma} e_\m^a = e^{-\sigma} \(  \tilde e ^ a _ \m + \p_\m y^a -  \om_\m^{ab} y_b + A_\m y^a\) ~, \quad \nabla_a\sigma = e^\sigma e^\mu_{a}\(\p_\mu\sigma+A_\mu\).
\ee
The diff- and Weyl-invariant measure is now given by $d^d x \det E = d^d x~ e^{-d\sigma}\det e$. 

As in the case of pure GR, one may consider a theory satisfying certain covariant constraints, analogous to the torsion-freedom condition of General Relativity. One such constraint is obtained by simply setting to zero the covariant derivative of the dilaton, $\nabla_a\sigma = 0$, which allows to eliminate the gauge field corresponding to dilations 
\be
\label{Amuconstr}
A_\mu = -\partial_\mu\sigma~.
\ee
Furthermore, generalizing the curvature two-form of eq.~\eqref{curvature2form} to the Weyl-invariant theory under consideration yields schematically: $\Omega^{-1} [D,D]\Omega \sim  T \hat P+ R \hat J+W\hat D$, where $W$---the field strength for the Abelian gauge field $A_\mu$, vanishes on the account of the constraint \eqref{Amuconstr}. The remaining two-forms, $T$ and $R$ are direct analogs of the torsion and curvature two-forms of GR, but now explicitly depending on the dilaton and its derivatives. In particular, imposing (the generalization of) the covariant torsion-freedom condition  $T=0$ allows to express the spin connection $\omega$ in terms of the vielbein and the dilaton (as well as their derivatives). The appropriately contracted curvature two-form $R$ then provides the standard two-derivative action for a Weyl weight-zero scalar, conformally coupled to gravity
\be
\int d ^ d x \det e~ e ^ {-(d-2)\sigma} \l [  \mathcal{R} +  (d-1)(d-2)(\p\sigma)^2 \r]~.
\ee
Higher-derivative Weyl-invariant terms can be obtained following the standard steps of the CCWZ construction, outlined above.

\section{CFT at large global charge: $U(1)$\label{sec_superfluid}}

We finally are in a position to discuss the coset construction for the symmetry breaking pattern of the conformal superfluid, described in the end of Section~\ref{general_strategy}:
\be
SO(d+1,1)\times U (1) \to SO(d)\times D'.
\label{U1_sbp}
\ee
For the purposes of realizing \eqref{U1_sbp}, we follow the previously outlined procedure to build a diffeomorphism-invariant theory with gauged Poincar\'e and Weyl symmetries, and subsequently freeze the d-bein $e_\mu^a$ to describe the non dynamical metric of our base manifold, the cylinder. The conformal group $SO(d+1,1)$ will emerge as the subgroup of the combined action of gauge symmetry and diffeomorphisms for which the non dynamical d-bein is left invariant.
It remains to decide which generators are spontaneously broken and which are not.
  According to the discussion around eq.~(\ref{local_conformal}) and in Appendix \ref{appB}, the generators acting in a local chart of the cylinder are naturally associated with those acting on the plane and obtained in the $R \to \infty$ limit. As emphasized in Section~\ref{general_strategy}, the unbroken group includes the effective time translations 
$\wh P _ 0' = \wh P _ 0 + \m \wh Q$, as well as the (Euclidean) spatial translations $\wh P _ i$ and rotations $\wh J _ {i j}$. The combined action of the latter local symmetries and of diffeomorphisms contains a local subgroup the isometry $SO(d)\times D'$ of the cylinder.
The  symmetry breaking pattern is thus, by construction, precisely that of eq.~(\ref{U1_sbp}).

It should be noted that our approach to realizing the symmetry breaking pattern (\ref{U1_sbp}) is not unique. One could imagine building a theory without bothering to gauge any symmetries, and focussing directly on the global ones (thus disposing of the diffeomorphisms). We are planning to investigate this possibility in the future, but for the present we choose to adopt a more redundant approach described above. We thus employ the coset construction for the following symmetry breaking pattern\footnote{At this stage we choose to work with the Minkowski signature on the cylinder.}
\be \begin {aligned} \text{broken:} \quad & \wh B _ i \equiv \wh J _ {0 i},~ \wh D, ~\wh Q,~  \\
\text{unbroken:} \quad & \wh P _ a ' = \wh P _ a + \m \, \d _ a ^ 0 \wh Q,~ \wh J _ {i j},
\label{eq:symmbreakpatt}
\end{aligned}
\ee
It is convenient to choose the coset representative in the following form
\be
\label{omega}
\Omega = e ^ {i y ^ a \hat P' _ a } e ^ {i \s \hat D} e ^ {i \eta ^ i \hat B _ i} e ^ {i \pi \hat Q} = 
e ^ {i  y ^ a \hat P _ a}  e ^ {i \s \hat D} e ^ {i \eta ^ i \hat B _ i} e ^ {i \chi \hat Q}, ~~~ \chi = \m t + \pi ~.
\ee
The parameter $\m$ will eventually be dynamically determined in terms of the charge $Q$ and the radius of the sphere $R$. 

Introducing, as in the above example of dilaton gravity, the appropriate gauge fields $\tilde e _ \m$, $\omega _ \m$ and $A_ \m$, the covariant derivative becomes
\be
 D _ \m = \p _ \m + i \tilde e _ \m ^ a \wh P _ a + \f {i} {2} \omega _ \m ^ {ab} \wh J _ {ab} 
+ i A _ \m \wh D~,
\ee
and the corresponding  Maurer-Cartan form reads
\bea
\label{eq:mc1f}
\Omega ^ {-1} D _ \m \Omega 
= i E_\m^b \l( \wh P_b ' + \nabla_b \pi \, \wh Q + \nabla_b \s \wh D+ \nabla_b \eta^i \wh B_i + \f{1}{2} \Xi_b^{ij}\wh J_{ij} \r) ,
\eea
where
\be
E_\m^b = e ^ {- \s} \Lambda ^ {~b} _ a e_\m ^a, ~~~ \nabla _ b \pi = e ^ \s \Lambda ^ c _ {~b} e ^ \n _ c \p _ \n \chi - \m \d ^ 0 _ b, ~~~ 
\nabla _ b \s = e^\s e^\n_{~d} \Lambda^d_{~b} \l( \p_\n \s + A_\nu \r).
\ee
Here $\Lambda_a^{~b}$ is a Lorentz transformation matrix, consisting solely of the boost Goldstones.\footnote{In terms of the velocity $\beta^{i}= \f{\eta^{i}}{\eta} \text{tanh}\, \eta$, the explicit components are: $\Lambda^{0}_{~0}= \gamma, \, \Lambda^{0}_{~i}= \gamma \beta_{i}, \, \Lambda^{i}_{~0}=\gamma \beta^{i}, \,  \Lambda^{i}_{j} = \delta^{i}_{~j} + (\gamma-1)\beta^{i} \beta_{j}/\beta^2 $.}
Expressions for the covariant derivative $\nabla \eta$ and connection $\Xi$ are not relevant for what follows, and we do not present them here. As before, the fields $e _ \m ^ a = \tilde e_\m ^ a + \p _ \m y ^ a - \omega _ {\m b} ^a y ^ b+A_\mu y^a$ and $\omega _ \m ^ {a b}$ are interpreted as the vielbein and the spin connection, and the corresponding field strengths (curvatures) are defined as follows
\bea 
\label{eqtwoform}
\Om^{-1} [ D_\mu,  D_\nu]\Om \equiv i E_\m^e E_\n^f  \l(T_{ef}^c P_c +  R_{ef}^{cd} J_{cd} + W_{ef} D \right) ~.
\eea
When written in terms of $e$ and $\omega$, the three 2-forms on the rhs of \eqref{eqtwoform} read
\begin{align}
T_{ef}^c&= e ^ {\s} \Lambda ^ {~c} _ a e ^ \m _ g e ^ \n _ h \Lambda ^ {g} _ {~e} \Lambda ^ {h} _ {~f} \l (  \p_\mu e_\nu^{~a} - \p_\n e_{\m}^{~a} - e_\m^{~a} A_\nu + e_\nu^a A_\m - e_{\m b} \om_\n^{ba} + e_{\nu b} \om_\mu^{ba} \r )~, \label{eq:FST_e}\\
R_{ef}^{cd}&= e ^ {2 \s} \Lambda ^ {~c} _ a \Lambda ^ {~d} _ b  e ^ \m _ g e ^ \n _ h \Lambda ^ {g} _ {~e} \Lambda ^ {h} _ {~f} 
\l ( \p_\m {\om}_\n^{ab} - \p_\n {\om}_\m^{ab} - \om _{\m c}^{a} \om_{\n}^{cb} + \om _{\n c}^{a} \om_{\m}^{cb} \r )~, \label{eq:FST_om} \\
W_{ef}&= e ^ {2 \s} \Lambda ^ {g} _ {~e} \Lambda ^ {h} _ {~f} e^{\m}_{~g} e^{\n}_{~h} \l ( \p_\m A_\n - \p_\n A_\m \r ) ~. \label{eq:FST_A}
\end{align}
These complete the list of the building blocks, necessary to write down the leading-order invariant Lagrangian. Any operator constructed from the covariant derivatives $\nabla \pi$, $\nabla \s$, $\nabla \eta$, the connection $\Xi$ and the field strengths in \eqref{eq:FST_e}-\eqref{eq:FST_A} in a way that respects the residual symmetry will be automatically invariant under the full local symmetry group.\footnote{Operationally, the whole procedure boils down to contracting the local Lorentz indices in an $SO(d-1)$ invariant way.}

\subsection{Constraints and the leading order Lagrangian}
\label{constrs}

The crucial difference between spontaneously broken space-time and internal symmetries is that for the former the number of Goldstone modes is usually smaller than the number of broken generators~\cite{Volkov:1973vd,Ivanov:1975zq,Low:2001bw,Delacretaz:2014oxa}. The phenomenon can be interpreted as if some of the Goldstone fields become massive, and therefore not visible from the low energy perspective. The way to implement this feature in the coset construction is via imposing covariant (inverse Higgs) constraints, that allow to express the would-be massive Goldstone fields in terms of the rest. (As emphasized above, one example of such a constraint is the standard torsion-freedom constraint of general relativity.)

We are honing in on describing the system with large charge under the simplest possible condition that the broken symmetries are non-linearly realized through the smallest possible number of low-energy fields. This requires imposing the full possible set of inverse Higgs constraints. 
Inspecting the transformation properties of various covariant derivatives, one can see that the following set of constraints is consistent with the underlying symmetry
\be
\nabla _ i \pi  =  0, ~~~ \nabla _ 0 \pi  =  0, ~~~ \nabla _ b \s = 0, ~~~  T_{bc}^{a}=0~.
\ee
(The last constraint should by now be familiar as the generalization of the standard standard torsion-freedom condition of GR)
These can be straightforwardly solved, the result being
\bea
\label{IH_solutions}
\b ^ i & = & \f {e ^ \m _ i \p _ \m \c} {e ^ \n _ 0 \p _ \n \c}, \qquad \m e ^ {-\s} = (e ^ \m _ a e ^ {\n a} \p _ \m \chi \p _ \n \c)^{1/2}, \qquad
A_\m = - \p_\m \s, \nn\\
\omega_\m ^{ab} & = & \f{1}{2} \bigg[ e^{\n a} \l(\p_\m e_\n^b-\p_\n e_\m^b\r) - e^{\n b} \l(\p_\m e_\n^a-\p_\n e_\m^a\r) \\ &-& e_{\m c}e^{\n a}e^{\lambda b} \l(\p_\n e^{\lambda c}-\p_\lambda e_\n^c\r)\bigg] - \l(e_\n^a e_\m^b-e_\n^b e_\m^a \r)A^\nu.   \nn
\eea
Upon imposing the inverse Higgs constraints only $\c$ and $e ^a _ \m$ remain as independent fields, while all the rest are algebraically expressed in terms of these. Moreover, at the leading order in the derivative expansion, only $E _ \m ^ a$ and $R _ {ef} ^ {cd}$ need be used as covariant building blocks.\footnote{Other covariant objects such as $\nabla _ a \eta ^ i$ and $\Xi _ a ^ {ij}$ (both functions of 
$\p _ \m \chi$ and $e _ \m ^ a$) are not needed at this order for they  only generate higher derivative terms in the Lagrangian.} 

We are now in a position to write down the simplest term in the action consistent with the desired symmetry breaking pattern. This is the invariant measure, which, according to the CCWZ prescription, reads 
\be 
\label{eq:leadingop}
{\mu}^{d}\det E = \det e \l (\p ^ \m \chi \p _ \m \c \r ) ^ {d/2}.
\ee
Furthermore, the two independent $SO (d-1)$ invariant contractions of $W _ {cd} ^ {ab}$ are (up to a total derivative)
\bea
\m ^ {-2} R_{ef}^{ef} &=& \f{\mathcal{R}}{\l| \p\c \r| ^ 2} -(d-1)(d-2)\f{ \nabla_\m \l| \p \chi \r|\nabla^\m \l| \p \chi \r|}{\l| \p \chi \r|^4}  - 2(d-1) \nabla_\m\(\f{\nabla^\m \l| \p \chi \r|}{\l| \p \chi \r|^3}\)~,\label{eqWcontr} \nn \\
\m ^ {-2} R_{0f}^{0f} &=&  \mathcal{R}_{\m \n} \f{\p^\m \c \p^\n \c}{\l| \p\c \r| ^ 4}~.
\label{W_leading}
\eea
Here $\mathcal{R}$ and $\mathcal{R}_{\m \n}$ denote respectively the ($d$-dimensional) Ricci scalar and the Ricci tensor, while $\nabla_\mu$ is the usual (metric-compatible) covariant derivative. 

At the leading order in the derivative expansion, the most general diff$\times$Weyl- invariant action thus is
\be \begin{aligned}
S
 = \,&  \frac{c_1}{6}  \int d ^ n x \det e \l| \p\c \r| ^ n  \\
+ \,&c _ 2  \int d ^ n x \det e \l| \p\c \r| ^ {(n-2)} \l [  \mathcal{R} +  (n-1)(n-2)\f{ \nabla_\m \l| \p \chi \r|\nabla^\m \l| \p \chi \r|}{\l| \p \chi \r|^2} \r] \\
+ \,&c _ 3  \int d ^ n x \det e \l| \p\c \r| ^ {(n-2)} \l [  \mathcal{R}_{\m \n} \f{\p^\m \c \p^\n \c}{\l| \p\c \r| ^ 2} \r]
 + \cdots~.
 \end{aligned}  \label{eq:actionU1}
\ee
The Wilson coefficients $c_i$ are the input parameters and cannot be derived from the EFT perspective -- they are determined by the specific underlying CFT. The loop expansion is governed by the parameter $\a (c _ i) E/\m$, where $E$ is a typical energy scale of the process under consideration and the constant $\a(c_i)$ is determined by the Wilson coefficients.
In the simplest case when the system becomes strongly coupled at $E \sim \mu$, naive dimensional analysis~\cite{Manohar:1983md,Georgi:1992dw, Cohen:1997rt} sets $c_i$ to be given by inverse powers of 
$4 \pi$. On the other hand, for a weakly coupled theory or a theory with an analog of large-$N$ dynamics the generic expectation is that $c _ i \gg 1$. At any rate, the coefficients $c_i$ are $\m$-, and hence, $Q$-independent, so that in the limit $Q \to \infty$ they can be effectively treated as $\mc O (1)$ parameters.

\subsection{Operators with the lowest dimension in $d=2+1$}
\label{u1}

After continuing to the Euclidean signature (where the Goldstone becomes $\c = - i \m \tau + \pi$), the generalization of the amplitude (\ref{GoldstoneAmplitude}) 
for a $d=2+1$ - dimensional CFT with an internal $U(1)$ symmetry becomes
\be 
\label{amplitude}
\langle Q|~ e^{- \wh H T} ~|Q \rangle = \int \mathcal{D}\chi ~e^{- \int d^3 x \det e\big[\mathcal{L}+i \frac{Q}{4 \pi R^2}\dot\chi  \big]}~.
\ee
Here $\mc L$ is the Euclidean analogue of the action (\ref{eq:actionU1})
\be
\mc L = -\f {c_1}{6} \l [ - \p _ \m \c \p ^ \m \c \r ] ^ {3/2} + \dots~ .
\label{euclidean_action_leading}
\ee
In the semiclassical approximation, the path integral is dominated by the saddle-point trajectory
$\dot\chi = - i\mu$, and (just as for a rigid rotor) the second term in the exponent under the integral in (\ref{amplitude}) fixes the value of $\m$ in terms of the charge
\be
\f {Q} {\mathrm{4 \pi}} = i R ^ 2 \frac{\partial{\mathcal{L}}}{\p\dot\chi} \Big | _ {\c = -i \m \tau} = 
\f {c _ 1} {2}(R\mu)^{2}+c_2+\mathcal{O}\((R\mu)^{-2}\)~,
\ee
This equation can be (perturbatively) inverted to solve for the parameter $\mu$
\be
R \m = \sqrt{\f{Q} {2 \pi c _ 1}} \l[ 1 - \f{2 \pi c _ 2}{ Q} +\cdots \r]~,
\label{mu_charge_rel}
\ee
so that $\mu\propto Q^{1/2}$ for $Q\gg 1$. Computing the corresponding action, one finds the lowest dimension in the sector with charge $Q$
\be
\label{eqdeltaexpansion}
{\Delta_Q} = \f {2} {3}\, \f {Q^{3/2}}{\sqrt{2 \pi c _ 1}} + 8 \pi c _ 2 \sqrt{\f{Q}{2 \pi c _ 1}}  +\mathcal{O}\(Q^{-1/2}\)~.
\ee

Note, in particular, that there is no contribution from the local Lagrangian that scales like the zeroth power of the charge. The $Q^0$ piece does however arise from the quantum corrections to the saddle point action. To evaluate this correction, consider the fluctuations around the semiclassical trajectory, $\c = -i\mu \tau + \pi$. Expanding the leading low-energy effective action \eqref{euclidean_action_leading} in $\pi$ and then canonically normalizing yields at the quadratic order
\be
\label{eqspi}
S^\pi_E =  - \f{1} {2} \int d^3 x \det e~ \pi \(\p_\tau^2 + \f{1}{2} \Delta_{S^2} \)\pi~.
\ee
Here $\Delta_{S^2} =- g^{ij}\nabla_i\nabla_j$ is the laplacian on the sphere (with eigenvalues $l (l+1) / R ^ 2$, $l =0,1,2,\dots$) and we have denoted the canonically normalized $\pi$ by the same symbol for notational simplicity. Notice that the speed of sound for the Goldstone fluctuations is fully model-independent, determined solely by the underlying symmetries. This is simply the consequence of conformal invariance: in $2+1$ dimensions, the tracelessness of the stress tensor for a perfect fluid with energy density $\rho$ and pressure $p$ (of which the theory described by \eqref{eqspi} is an example) requires $c_s^2 = dp/d\rho = 1/2$. 

The dispersion relation for the Goldstone fluctuations is
\be
\label{omega_energy}
\omega_\ell =\frac{1}{R}\sqrt{ \frac{\ell (\ell+1)}{2}}~,
\ee
and the energies of the excited states featuring $n_\ell$ modes of angular momentum $\ell$ are simply given by the sum 
\be
E _ Q ^{(n_1,\dots)} R = \Delta_Q+\sum_\ell n_\ell \omega_\ell.
\label{excited}
\ee
Notice that $\omega_1=1$, so that acting $n$ times on the ground state with the creation operator $a^\dagger_1$ for the $\ell=1$ modes generates its descendant with scaling dimension $\Delta_Q+n$. In contrast, acting with powers of $a^\dagger_\ell$ with $\ell\neq 1$ generates other primaries, including those of higher spin.

Computing the leading order (one-loop) quantum correction to the 1-PI action amounts to evaluating the following functional determinant
\be
\Gamma_{\text{1-loop}}  = \f{1}{2} \ln  \det \l[ -\p_\tau^2 - \f{1}{2} \Delta_{S^2} \r]~.
\label{eq:GammaU1}
\ee 
The calculation is straightforward, but not without subtleties. The details are spelled out in Appendix \ref{app}, and we will just quote the result for the quantum-corrected version of Eq. \eqref{eqdeltaexpansion}:
\be
\Delta_Q = \f {2} {3}\, \f {Q^{3/2}}{\sqrt{2 \pi c _ 1}} + 8 \pi c _ 2 \sqrt{\f{Q}{2 \pi c _ 1}} - 0.0937256 +\mathcal{O}\(Q^{-1/2}\)~.
\label{eqdelta0}
\ee
The third term on the rhs of this equation is a true prediction of the theory: no local counterterm can renormalize it, since the local EFT \eqref{eq:actionU1} does not contain operators that scale as $Q^0$ when evaluated on the background solution.

\section{CFT at large global charge: $U(1)\times U(1)$}
\label{u1timesu1}

An interesting question is how things change for more complicated internal symmetries, different from a simple $U(1)$. In this section we set out with exploring the next-to-simplest case of a CFT with a $U(1)\times U(1)$ symmetry, focussing on the sector with non-zero charges $Q_1$ and $Q_2$ (corresponding to each of the two Abelian factors). 

It is instructive to first look at a simple example which nicely illustrates some of the subtle aspects of the general construction. To this end, consider a 4d (Minkowskian) classical CFT featuring two complex scalars $\Phi_1$ and $\Phi_2$ with charges $(1,0)$ and $(0,1)$ under the two groups
\be
\label{example}
\mathcal{L}=|\p\Phi_1|^2+|\p\Phi_2|-\frac{\lambda_1}{4}|\Phi_1|^4-\frac{\lambda_2}{4}|\Phi_2|^4-\frac{\lambda_{12}}{2}|\Phi_1|^2|\Phi_1|^2~.
\ee
In a state with both $U(1)$ charges non-zero and large, one generically expects that the vevs of both scalars are non-vanishing, so that they can be parametrized in terms of the radial modes and phases
\be
\Phi _{i}= \frac{a_{i}}{\sqrt{2}}~e^{i\chi_{i}}~,
\ee
where the index $i={1,2}$ runs over the two $U(1)$ groups. As before, projecting onto the appropriate state with non-zero $Q_i$ amounts to adding the  operator $-\sum_i\dot\chi_i Q_i/{\rm Vol}$ to the Lagrangian.~Requiring then that the Lagrangian is stationary with respect to variations of fields at boundaries fixes the two charge densities as
\be
\label{charges12}
\rho_{i}\equiv \frac{Q_{i}}{\rm Vol}=  a^2_{i}\dot\chi_i ~.
\ee
Just as for a rigid rotor, non-zero values of the charge densities provide centrifugal forces that keep the radial modes' vevs away from zero, and to find the latter one has to minimize the following effective potential
\be
\label{veff}
V_{\rm eff} = \frac{\rho_1^2}{2 a_1^2} + \frac{\rho_2^2}{2 a_2^2}+\frac{\lambda_1}{4} a_1^4+\frac{\lambda_2}{4} a_2^4 +\frac{\lambda_{12}}{2} a_1^2 a_2^2~.
\ee
For a generic state with both charges non-vanishing, $a_{1,2}\neq 0$ and the internal group is fully broken; however, for special cases one may have a partial symmetry restoration. As an interesting example one can consider the limit where one of the charges, e.g. $Q_2$, is sent to zero. It is straightforward to show that for a \textit{positive} $\lambda_{12}$, the minimum of $V_{\rm eff}$ corresponds to a vanishing $a_2$ in this limit,\footnote{We will always assume positive $\lambda_1$ and $\lambda_2$ so that the full potential in \eqref{example} is bounded from below.} so that the corresponding $U(1)$ group is restored.
For a \textit{negative} $\lambda_{12}$, on the other hand,\footnote{One can always choose $\lambda_{12}$ to be negative without compromising stability of the full theory \eqref{example}, as far as its magnitude is small enough.} $a_2=0$ no longer minimizes $V_{\rm eff}$, as can be easily seen by noting that the effective ``mass squared", $m_2^2= \lambda_{12} a_1^2$, of $a_2$-fluctuations in \eqref{veff} is negative. The second $U(1)$ thus remains broken even in the limit of vanishing $Q_2$. 

The above discussion straightforwardly generalizes to theories that feature more fields, possibly carrying complex charge assignments under the internal symmetry. In that case, one expects a qualitatively similar structure in the space spanned by the charges: a generic point will correspond to a fully broken symmetry, while there may be directions along which the symmetry is (partially) restored. For all charges non-vanishing, the low-energy limit of the system is generically described by a theory of Goldstone bosons $\chi_i$ that acquire vacuum expectation values $\dot\chi_i=\mu_i$ fixed by the corresponding charge densities. It then follows from the above discussion that depending on the details of the UV theory, the Goldstone description may or may not break down along certain directions (analogous to $\mu_2\to 0$ in the simple example of eq.~\eqref{example}).~This highlights the general pattern that emerges when dealing with a low-energy description of systems in a state with multiple global charges: a state belonging to a generic point in $\mu$-space will be amenable to a low-energy description in terms of Goldstone bosons; for states that belong to certain special directions in that space, however, such a description may fail due to a (partially) restored symmetry. 

For the purpose of studying the most general low-energy CFT at large $U(1)\times U(1)$ quantum numbers, one can straightforwardly generalize the coset construction of the previous section. Just as before, it is possible to impose the inverse Higgs constraints (on an arbitrary linear combination of $\nabla\pi_1$ and $\nabla\pi_2$) to eliminate the Goldstone modes associated with the dilatations and the Lorentz boosts. As a result, the low-energy dynamics now features a pair of Goldstones corresponding to each of the broken internal symmetries. Likewise, the most general action consistent with the desired symmetry breaking pattern can be constructed in complete analogy with the case of a single internal $U(1)$; there will be a leading set of operators in the derivative expansion (cf. eq.~\eqref{eq:leadingop}), supplemented by operators suppressed by powers of $\p/\mu$ or $1/(R\mu)^2$. 

One important difference is that in contrast to the case of a single $U(1)$, there is now a functional freedom in writing the most general leading-order action
\be
\label{equ1timesu1}
S= \int d^3 x ~\det e~ |\p\chi_1|^{3/2}|\p\chi_2|^{3/2}~P\(\frac{\p\chi_1\cdot\p\chi_2}{|\p\chi_1||\p\chi_2|},\frac{|\p\chi_2|}{|\p\chi_1|}\)~,
\ee
where $X\equiv(\p\chi_1\cdot\p\chi_2)/(|\p\chi_1||\p\chi_2|)$ and $Y\equiv|\p\chi_2|/|\p\chi_1|$, and we have assumed a generic situation in which both $\chi_1$ and $\chi_2$ are in the superfluid phase with $\dot\chi_{1,2}=\mu_{1,2}\neq 0$.

The action \eqref{equ1timesu1} is clearly Weyl-invariant,\footnote{The quantities $X$ and $Y$ are the only two independent Weyl-invariant scalars one can write down at the given order in the derivative expansion.} and the two $U(1)$ symmetries are realized as symmetry under constant shifts of $\chi_1$ and $\chi_2$.
The (leading-order) expressions for the two charges are
\begin{align}
Q_1 = \frac{3}{2} r \mu^2 P \bigg(1 - \frac{2}{3} \frac{Y P_Y}{P}\bigg)~,~~
Q_2=\frac{3}{2r} \mu^2 P \bigg(1 + \frac{2}{3} \frac{Y P_Y}{P}\bigg)~,~~\frac{Q_1}{Q_2} = f(r)~,
\end{align}
where we have defined $\mu^2 \equiv \mu_1\mu_2$ and $r^2 \equiv \mu_2/\mu_1$.  The subscripts on $P$ denote differentiation with respect to the given argument, and $P$ and all its derivatives are assumed to be evaluated on the background solution with $X=1,~Y=r^2~$. One can see that the charges have a simple scaling with respect to a common rescaling of $\mu_1$ and $\mu_2$, that leaves the ratio $r^2$ unchanged. Upon scanning all possible values of $r$, however, one may encounter singularities in the function $P$, associated with directions that correspond to phases with (partially) restored internal symmetry.

Denoting $Q\equiv \sqrt{Q_1 Q_2}$, the same arguments that led to \eqref{eqdeltaexpansion} yield the following semiclassical result for the scaling dimension of the lowest-lying operator
\beq
\label{eqdeltau1xu1}
\Delta_{Q_1,Q_2}=\gamma_{3/2}\(r\)~Q^{3/2} + \gamma_{1/2}\(r\)~Q^{1/2}  +\mathcal{O}\(Q^{-1/2}\)~.
\eeq
Note, in particular, that the $Q^0$ contribution is absent at the semiclassical level -- just as it was in the case of a single $U(1)$. This contribution comes back, however, with the inclusion of quantum effects. To evaluate it, we consider small perturbations on the background at hand, $\pi_{1,2}=\mu^{-1}_{1,2}\(\chi_{1,2}-\mu_{1,2}t\)$. We will find it convenient to further define $\pi_{\pm}=(\pi_1\pm\pi_2)/2$ in terms of which the quadratic (Euclidean) action for fluctuations reads
\begin{align}
\label{eqpilagu1u1}
S^{\pi}_E &= 3 P\mu^3\int d^3 x ~\det e~ \bigg[ (\p_\tau\pi_+)^2-\frac{1}{2}\(\vec\nabla\pi_+\)^2-\frac{4}{3} \frac{YP_Y}{P}\(\p_\tau\pi_+ \p_\tau\pi_--\frac{1}{2}\vec\nabla\pi_+\vec\nabla\pi_-\)\nonumber \\
& + \(-\frac{1}{2}+\frac{2}{3}\frac{Y P_{Y}}{P}+\frac{2}{3}\frac{Y^2 P_{YY}}{P}\) (\p_\tau\pi_-)^2-\(\frac{1}{2}-\frac{2}{3}\frac{X P_{X}}{P}\)\(\vec\nabla\pi_-\)^2~.
\end{align}
Here the quantities $X, Y,P, P_X$, etc.~are understood as evaluated on the semiclassical background. Stability and subluminality of small fluctuations requires these to satisfy certain constraints. We will not reproduce these constraints here, but we note that the function $P$ can always be chosen such that they are all met. One can straightforwardly diagonalize the action \eqref{eqpilagu1u1} to find the propagation speeds for the two modes. One of these still propagates at half of the speed of light. The other mode, $\pi_-$, has the speed of sound $c_-$ which, depending on the precise form of the action, can lie anywhere between $0$ and $1$. 

The generalization of the expression for the one-loop quantum effective action to the case with the $U(1)\times U(1)$ internal symmetry reads
\be
\Gamma_{\text{1-loop}}  = \f{1}{2} \ln  \det \l[\( -\p_\tau^2 - \f{1}{2} \Delta_{S^2} \)\( -\p_\tau^2 - c^2_{-} \Delta_{S^2} \)\r]~.
\label{eq:GammaU1xU1}
\ee 
Note that unlike CFTs with a $U(1)$ global group, the $Q^0$ correction to Eq. \eqref{eqdeltau1xu1} is not a fixed number. However, ``universality" is still there to the extent that the equation of state of the low-energy fluid---that is, the function $P(X,Y)$---is known. 

The details of the calculation of the expression in eq. \eqref{eq:GammaU1xU1} are spelled out in Appendix \ref{app}, and the result reads
\be
\Delta_{Q_1,Q_2} = \gamma_{3/2}(r)~ Q^{3/2} + \gamma_{1/2}(r) ~Q^{1/2}-(1+\sqrt{2} c_{-})\cdot 0.0937256 +\mathcal{O}\(Q^{-1/2}\)~.
\label{eqdelta0u1xu1}
\ee
The precise value of the constant piece in $\Delta_0$ depends on a single number -- the speed of sound $c_-$ of the second Goldstone mode. Notice that while no more a fixed number, causality ($0\leq c_{s2}\leq 1$) constrains the coefficient $\gamma_0$ of the $Q^0$ contribution to lie in the range $-(1+\sqrt{2})\cdot 0.0937256 \leq \gamma_0 \leq - 0.0937256$.

Apart from defining the lowest dimension, the coefficient $c_-$ will also enter in the expression for the energy of excited states. For instance, states featuring two Goldstone bosons with the speeds of sound $1/\sqrt{2}$ and $c_-$ will have energy
\be
\D ^ {(l_+,l_-)} = \D _ {Q_1,Q_2} + \sqrt{\f{l_+(l_+ + 1)} {2}} + \sqrt{ c _ {-} ^ 2 {l_-(l_- + 1)}},
\ee
where $l _ {\pm}$ label the momentum modes of $\pi _ \pm$ respectively. Just as in the case of single $U(1)$, the $+$ mode with $\ell=1$ generates the descendants of the ground state ($\omega^+_1=R^{-1}$), while acting with the creation operators of the $-$ modes, as well as those of the $+$ modes with $\ell\neq 1$, gives rise to other primaries of various spin.


\section{CFT at large global charge: $SO(3)$ \label{sec_SO3}}
Our final example corresponds to the case of a non-Abelian internal $SO(3)$ group, whose charges $\wh Q_\a, \a=1,2,3$, satisfy the commutation relations $[\wh Q_\a, \wh Q_\b]= i \epsilon_{\a\b\g} \wh Q_\g$.  We consider an eigenstate of $\wh Q_3$, which, in line with our general discussion, we 
describe  by the coset for the symmetry breaking pattern
\be \begin {aligned} \text{broken generators :} \quad & \wh B _ i \equiv \hat J _ {0 i},~ \wh D, ~\wh Q_1, ~\wh Q_2, ~\wh Q_3,~  \nn \\
\text{unbroken generators :} \quad & \hat P' _ a = \wh P _ a + \m \d _ a ^ 0 \wh Q_3, ~ \wh J _ {i j}~.
\end{aligned}
\ee
We will parametrize the $G/H_0$ coset in the following way
\be
\Omega = e ^ {i \wh P' _ a y ^ a} e ^ {i \s \wh D} e ^ {i \eta ^ i \wh B _ i} e ^ {i \pi _ 3 \wh Q_3} e ^ {i \pi_I \wh Q_I} =  e ^ {i  \wh P _ a y ^ a} e ^ {i \s \wh D} e ^ {i \eta ^ i \wh B _ i} e ^ {i \chi \wh Q^3} e ^ {i \pi_I \wh Q_I}~, ~~~ \c =\m t + \pi _ 3.
\ee
Here and henceforth, the capital letter index $I$ exclusively denotes the indices $1,2$. It should be noted that the choice of the coset parametrization is in part dictated by requiring that $\wh Q _ 3$ be the conjugate momentum to $\pi _ 3$: it can be trivially seen that 
$\pi _ {1,2}$ do not transform under the action of $\wh J _ 3$, while $\pi _ 3$ shifts by a constant.

The vielbein and the covariant derivatives for the Goldstone fields are given by
\bea
E_\m^b &=& e ^ {- \s} \Lambda ^ {~b} _ a e_\m ^a, ~~~ \nabla _ b \s = e^\s e^\n_{~d} \Lambda^d_{~b} \l( \p_\n \s + A_\nu \r), \\
\nabla_b \pi _ 3 &=& e^{\s} \Lambda^d_{~b} e^\n_{~d} \l[\p_\n \c \mathcal M_{33}  + (\mathcal M^{-1} \p_\n \mathcal{M})_{12} \r] - \m \delta_b^0, \\
\nabla_b \pi^J&=& e^{\s} \Lambda^d_{~b} e^\n_{~d} \l[\p_\n \c \mathcal M_{3J}  + (\mathcal M^{-1} \p_\n \mathcal{M})_{3I} \epsilon_{IJ} \r],
\eea
where the rotation matrix $\mathcal M$ is defined as $(\mathcal M)_{\a\b} \equiv \l(e^{i \pi ^ I Q_I}\r)_{\a\b}$. As for the previously explored examples with $U(1)$ internal symmetries, it is possible to impose the inverse Higgs constraints 
\be
\nabla _ i \pi _ 3  =  0, ~~~ \nabla _ 0 \pi _ 3  =  0, ~~~ \nabla _ b \s = 0~,
\ee
which will reduce the field content of the low-energy theory to just $\pi_1$, $\pi_2$ and $\pi_3$.
Introducing the following notation 
\be
\c _ a = e ^ \m _ a \l [ \p_\m \c \mathcal M_{33}  + (\mathcal M^{-1} \p_\m \mathcal{M})_{12} \r ], ~~~
\pi ^ J _ a = e ^ \m _ a \l[\p_\m \c \mathcal M_{3J}  + (\mathcal M^{-1} \p_\m \mathcal{M})_{3I} \epsilon_{IJ} \r]
\label{chi_pi_def}
\ee
the Goldstone modes associated with  Lorentz boosts and dilations and the gauge field $A_\mu$ can then be written as
\be
\b ^ i = \f {\c _ i} {\c _ 0},
~~~~ \m e ^ {-\s} = \sqrt {\c ^ a \c _ a } \equiv \sqrt {\c \cdot \c }, ~~~~A_\m = - \p_\m \s .
\label{SO3_IH_solutions}
\ee
The remaining building blocks of the invariant action are
\bea
\m ^ {-1} \nabla _ 0 \pi ^ J & \equiv & X ^ I = \f {\c \cdot \pi ^ J} {\c \cdot \c }, ~~
\nabla _ i \pi ^ J = \f {\m} {\sqrt {\c \cdot \c }} \l [ \pi _ i ^ J - \f {\c _ i \pi ^ j _ 0} {\sqrt {\c \cdot \c }} + \f{( \vec \pi^J \vec \c ) \c _ i} {\vec \c ^ 2} 
\l ( \f {\c _ 0} {\sqrt {\c \cdot \c }} - 1 \r ) \r ], \nn \\
\nabla _ i \pi _ I \nabla _ i \pi _ J & \equiv & Y ^ {I J} = \f {(\c \cdot \pi ^ I) (\c \cdot \pi ^ J)} {\c \cdot \c} - \pi ^ I \cdot \pi ^ J. \nn
\eea
Hence, the leading-order action is written as
\be
S= \f {c} {6} \int d^3 x ~\det e~ \l ( \c \cdot \c \r ) ^ {3/2}~P\(X ^ I, Y ^ {I J}\)~.
\ee
The general solution of the equations of motion for a configuration with fixed charge is
\be
\pi _ 3 = \c _ 0, ~~~~ \pi ^ I = v ^ I (P),
\ee
where $v^I (P)$ are $Q$-independent constants, determined by the form of the function $P$ and $\c _ 0$ is an arbitrary constant (the analog of $\vp _ 0$ discussed at length in Section \ref{sec_rotor}). The spectrum can be found by expanding the action up to quadratic order in $\pi_1$, $\pi_2$ and $\pi_3$ around this solution. A similar situation, albeit without the conformal symmetry, was considered in~\cite{Nicolis:2013sga}. It was observed that for a special case, which corresponds to 
\be
\f {\p P} {\p X ^ I} \bigg | _ {\pi ^ I = v ^ I} = 0~,
\ee
there are two modes in the spectrum with fixed masses, zero and $\m$ correspondingly, while the mass of the third mode is theory-dependent (i.e. it depends on the various Wilson coefficients of the low-energy EFT). Introducing small but non-vanishing $b _ I = {\p P} / {\p X ^ I}$ leaves the zero mode intact while it changes mass of other two modes by terms of order $\sim b_I ^ 2$. Therefore, unless the coefficients in the Lagrangian are tuned in such a way as to make the masses of the two massive Goldstones small, there is only one zero mode and the low energy dynamics is described by only one Goldstone $\pi_3$.

The same argument used for the case of a single $U(1)$ internal group then yields the following semi-classical result for the scaling dimension of the lowest-lying operator
\be
\label{eqdeltaexpansionSO3}
{\Delta} =\gamma_{3/2} ~Q_3^{3/2} + \gamma_{1/2}~Q_3^{1/2}  +\mathcal{O}\(Q_3^{-1/2}\)~,
\ee
where $\gamma$'s are constants. Moreover, given the similarity of the infrared physics, the quantum-corrected version of the lowest scaling dimension is fully identical to (\ref{eqdelta0}). 

For a non-Abelian group of rank $N$ the situation will be similar. In general, provided the whole group is broken down spontaneously, only the Goldstones corresponding to the Cartan generators will be massless. In this case certain quantities, such as the lowest dimension, will be independent of the specific choice of the group. In particular they will coincide with the ones derived for the $U (1)_1 \times U (1) \times \dots \times U(1)_N$ case. However, as we discuss in the end of the next section, the way various operators are matched in the IR onto expressions in terms of the Goldstone bosons will be sensitive to the global structure of the group.


\section{$n$-point functions}
\label{npt}
The methodology described above can be readily applied to access CFT data beyond the operator spectrum. As an obvious generalization of the previous analysis one can evaluate, on the cylinder, $n$-point functions of local operators, such as the conserved currents or the stress tensor, between the $in$ and $out$ states with large internal charge.
Upon mapping onto the plane, this will describe $(n+2)$-point functions with an insertion of the lowest-lying operator of charge $Q$ both at the origin and at infinity. 
We believe the issue of  $n$-point functions warrants a more detailed analysis. We leave that for future work and we here   provide the basic  remarks, focussing mostly on the case of a single $U(1)$.

Let us start by recalling that in the limit where the $in$ and $out$ states are well-separated in the cylinder time, $\tau _ {out} - \tau _ {in} \equiv T \to \infty$, the path integral in (\ref{GoldstoneAmplitude}) evaluates to
\be
\la Q | e ^ {-(\tau _ {out} - \tau _ {in}) \wh H} | Q \ra =  \, e ^ {- \D _ Q (\tau _ {out} - \tau _ {in})/R}~.
\label{2pt_cyl}
\ee
(Notice that any---in general $Q$-dependent---prefactor that may appear in \eqref{2pt_cyl} can always be rescaled away by suitably normalizing the state $| {Q} \ra$. We have assumed such a normalization above.) It can be easily checked that, upon using the map between correlators on the cylinder and those on the plane, $\la O(x)\dots\ra_{e^{2\sigma(x)}dx^2}=e^{-\sigma\Delta_{\mathcal{O}}} \la O(x)\dots\ra_{\mathbb{R}^d} $, eq.~(\ref{2pt_cyl}) results  as the leading term in the expansion of the standard result for the two-point function
\be
\la \mc O _ {-Q} (x_ {out}) \mc O _ {Q} (x_ {in}) \ra = \f {1} {(x_ {out} - x_ {in} ) ^ {2 \D _ Q}}~.
\ee
In what follows, we list some of the other (simplest) relevant correlators  that can be accessed through the semiclassical analysis in a CFT with a global $U(1)$ symmetry. 

\subsection{Three-point function with an insertion of the $U(1)$ current} 

Perhaps the next simplest observable one can calculate in the low-energy theory is the three-point function with an insertion of the $U(1)$ current:
\be
\label{plane3pt}
\la \mc O _ {-Q} (x_ {out}) j _ 0 (x) \mc O _ {Q} (x_ {in}) \ra.
\ee
Using the low-energy expression for $j_0$ in terms of the Goldstone degrees of freedom, \eqref{plane3pt} can be readily evaluated as it is directly related to the following expectation value on the cylinder
\be
\la Q, \tau _ {out} | i \f {\p \mc L}{\p \dot \c} (\tau, \mathbf n) | Q, \tau _ {in} \ra = \f {Q} {4 \pi R ^ 2} \, e ^ {- \D _ Q (\tau _ {out} - \tau _ {in})/R}~.
\ee
It is important to note that the charge $Q$ is not renormalized, so that the tree level result is exact. Transforming the cylinder correlator onto the plane (including the appropriate Jacobian factors), we find
\be
\la \mc O _ {-Q} (x_ {out}) j _ 0 (x) \mc O _ {Q} (x_ {in}) \ra 
\underset{
\ba{lll}
\scriptstyle x_{out} \to \infty \\
\scriptstyle x_{in} \to 0
\ea
}
{=} \f {Q}{4 \pi}~ \f{1}{ | x _ {out} |^{2 \D _ 0} |x|^ 2} ~.
\ee
In the relevant limit $r_ {out}\to \infty$ and $r_ {in} \to 0$, this is in perfect agreement with the standard expression for the CFT tree-point function of two scalars and a vector (see, e.g., \cite{Rychkov:2016iqz}).

\subsection{Three-point function of charged scalars} 

A further example we wish to consider is  the three-point function of charged scalar primaries 
\be
\label{3pt}
\la {\cal O}_{- (Q+q)}(x_{out}){\cal O}_{q}(x){\cal O}_{Q}(x_{in}) \ra~,
\ee
where we assume $q\ll Q$. 

According to the general strategy discussed in Section~\ref{CCWZ}, an operator transforming linearly under the broken group can be reconstructed in terms of the Goldstone fields. This amounts to finding, by matching the quantum numbers, an appropriate representation of $\mathcal{G}$, that contains the representation of the unbroken group. For the symmetry breaking pattern at hand, eq.~\eqref{eq:symmbreakpatt}, the representations of the unbroken group are generated by rotations ($J _ {12} \equiv J _ 3$), and so are labeled by an integer $n$ (spin)
\be
\phi ' \equiv \rho (e ^ {i J _ {12} \a} ) \phi = e ^ {i n \a} \phi.
\ee
Therefore, for any representation $\kappa$ of $\mathcal{G}$ that contains the spin $n$ subrepresentation, the field (following the same notation as in  eqs.~\eqref{reps1} and \eqref{reps2})
\be
\Phi \equiv \kappa (e ^ {i \s \wh D} e ^ {i \wh B _ i \eta ^ i} e ^ {i \c \wh Q}) \tilde \phi
\ee
transforms linearly. We are interested in scalar operators with definite scaling dimension $\d$ and charge $q$, so we choose the representation $\kappa$ characterized by these quantum numbers. Using the expression (\ref{IH_solutions}) for $\s$ in terms of $\c$, the operator $\mc O _ q$ becomes
\be
\label{IRop}
{\cal O}_{q, \d} = C \l ( \p \c \r ) ^ {\d} e ^ {i \c q} + \dots~.
\ee
where $C$ is an incalculable constant that depends on the operator
${\cal O}_{q}$ and on the underlying theory. The ellipses refer to extra, curvature-dependent contributions that have the correct quantum numbers to enter the expression for the interpolating operator. These contributions are suppressed by factors of $Q^{-1}$ and will not play any important role in the discussion to come, so we will discard them from now on. 

The cylinder counterpart of the three-point function (\ref{3pt}), 
\be
\label{3ptcyl}
\la Q + q,\tau _ {out}| {\cal O}_{q, \d}(x) | Q ,\tau _ {in}\ra~,
\ee 
can be evaluated by slightly modifying the path integral in a way that accounts for unequal charges of the $in$ and $out$ states. This casts eq.~\eqref{3ptcyl} into the following form
\bea
\label{integral}
 C \int d \c _ i \, d \c _ f \,  \int _ {\c _ i, \c_f} \mc D\c ~(\partial\chi)^\delta  \exp \bigg[  -S\l [\c \r]  - \f{iQ}{4 \pi R ^ 2} \int d\tau d^{2}{\bf n} \, \dot \c + i q (\c(x)-\c _ f) \bigg ].
\eea
The above integral can be computed in the saddle point approximation around a semiclassical configuration with charge $Q$. This immediately yields the leading order result
\be
\label{3pt1}
\la Q + q,\tau _ {out}| {\cal O}_{q, \d}(x) | Q ,\tau _ {in}\ra = C_q \mu ^ \d \, e ^{\m q (\tau-\tau _ {out})} e ^ {- \D _ Q (\tau _ {out} - \tau _ {in})/R}.
\ee
Alternatively, we could have chosen to compute \eqref{integral} around a different saddle point configuration with charge $Q + q$:
\be
\label{3pt2}
\la Q + q,\tau _ {out}| {\cal O}_{q, \d}(x) | Q ,\tau _ {in}\ra = C_q \m ^ \d \, e ^{\m q (\tau-\tau _ {in})} e ^ {- \D _ {Q+q} (\tau _ {out} - \tau _ {in})/R}.
\ee
It suffices to use the relation (which follows from eqs.~(\ref{mu_charge_rel}) and (\ref{eqdeltaexpansion}))
\be
\label{chempot}
R\mu = \D _ {Q+1} - \D _ Q \approx \f {\p \D} {\p  Q},
\ee
to show the equivalence of the two representations \eqref{3pt1} and \eqref{3pt2}. (Notice that eq.~\eqref{chempot} is the standard statistical mechanics definition of the chemical potential.) 

One can straightforwardly check that upon mapping onto the plane, \eqref{3pt1} implies the correct form of the three-point function (taken in the appropriate limit $x_{1}\to \infty,~x_{3}\to 0$)
\be
\la\mathcal{O}_1(x_1)\mathcal{O}_2(x_2)\mathcal{O}_3(x_3)\ra = \frac{\lambda_{1,2,3}}{|x_{12}|^{2\alpha_{123}}|x_{13}|^{2\alpha_{132}}|x_{23}|^{2\alpha_{231}}}~,
\ee
with $x_{ij}=x_i-x_j$ and $\alpha_{ijk}=(\Delta_i+\Delta_j-\Delta_k)/2$~.~In particular, it follows from the semiclassical result \eqref{3pt1} that the three-point function of two large charge primaries  and a small charge primary of dimension $\delta$ satisfies at leading order in the $1/Q$ expansion the scaling law
\be
\lambda_{Q+q,q,Q}\propto \frac{C_q}{c_1^{\delta/2}}~ Q^{\delta/2}.
\label{struct_const}
\ee
Similarly we could consider three point functions involving the spinning operators associated to the small excitations around the ground state $|Q\rangle$ and $|Q+q\rangle$  whose dimension is dictated by eq.~(\ref{excited}).
The corresponding three-point function coefficients are controlled by the same parameters and by the same scaling as in eq.~(\ref{struct_const}). This will be clarified in the next section by the study of the 4-point function and of its OPE decomposition.

\subsection{Four-point function of charged scalars} 

The last correlator we wish to explore in this section is that of four charged scalar primaries 
\be
\label{4ptplane}
\la {\cal O}_{- Q}(x_{out}){\cal O}_{q_2, \d_2}(x_2){\cal O}_{q _ 1, \d_1}(x_1){\cal O}_{Q}(x_{in})\ra.
\ee
We will compute this correlator in two different regimes, corresponding to large and small separations $|x _ 2 - x _ 1|$ between the two insertions. The effective field theory approach is clearly not applicable for an arbitrarily small $|x _ 2 - x _ 1|$; in particular, it is expected that the EFT is only valid for separations between the two insertions on the cylinder that are larger than the inverse cutoff. The precise condition is
\be
\sqrt{ (\tau _ 2 - \tau _ 1 ) ^ 2 + \t ^ 2 R ^ 2} \gg \f {1} {\m} \sim \f {R}{\sqrt{Q}},
\ee
where $\sqrt{2}\t _{12} $ is the angular distance between $x_1$ and $x_2$ on the spatial sphere. 

For simplicity of presentation, we will use the following notation for the cylinder counterpart of \eqref{4ptplane}:
\be
F _ {q_1 q _ 2} ^ {\d _ 1 \d _ 2} = \la Q,\tau_{out}| {\cal O}_{q_2, \d_2}(\tau_2, {\bf n}_2){\cal O}_{q _ 1, \d_1}(\tau_1,{\bf n}_1)| Q, \tau_{in}\ra\, .
\label{4pt_OO}
\ee
With the representation \eqref{IRop} for the operators of interest, one can readily evaluate the semiclassical contribution to this quantity, along with its leading corrections in the $1/Q$ expansion:
\bea
\label{F}
\begin{split}
F _ {q_1, q_2} ^ {\d _ 1 \d _ 2} &= C_1 C _2 (\m) ^ {\d _ 1 + \d _ 2} e ^ {\m (q _ 1 \tau _ 1 + q _ 2 \tau _ 2)} e ^ {- \D _ Q (\tau _ {out} - \tau _ {in})/R} \\ &\times  \bigg ( 1 -  \bigg \la q_1 q_2 \pi_2\pi_1 +\frac{\delta_2 q_1}{\mu}\dot\pi_2\pi_1+\frac{\delta_1 q_2}{\mu}\pi_2\dot\pi_1 \bigg \ra \bigg )\int _ 0 ^ {2 \pi} e ^ {i (q _ 1 + q _2) \c _ 0} \f {d \c _0}{2 \pi} ~,
\end{split}
\eea
where we have denoted $\pi_i=\pi(\tau_i, {\bf n}_i)$. Notice that integrating over $\chi_0$ enforces charge conservation: $q_1=-q_2\equiv q$.

The various correlators of $\pi$ in \eqref{F} can be found by expanding the (canonically normalized) field fluctuations in terms of creation and annihilation operators of definite angular momentum:
\be
\pi (\tau,\mathbf n)= \pi _ 0 (\tau) + \sum_{\ell\neq 0,m} \frac{1}{\sqrt{2\omega_\ell}}\(Y_{\ell m}({\mathbf n}) a_{\ell m} e^{-\omega_\ell \tau}+Y^*_{\ell m} (\mathbf n)a^\dagger_{\ell m} e^{\omega_\ell \tau}\),
\ee
where we have explicitly separated the zero mode $\pi _ 0 (\tau)$ from the rest and $\omega _ \ell$ is defined in~(\ref{omega_energy}).

\subsubsection*{Large separations}

At separations larger than the radius of the spatial sphere, $\tau_2-\tau_1\gg R$, the modes with $\ell\geq 2$ exponentially decouple compared to the $\ell = 0$ and $\ell=1$ modes. As discussed above, the latter mode has energy $\omega_1=R^{-1}$, so it generates the descendants of the ground state.
Neglecting the $\ell\geq 2$ modes, and taking into account that the correlator of the zero mode is given by
\be
\la \pi _ 0 (\tau _ 2) \pi _ 0 (\tau _ 1) \ra = - \f {|\tau _ 2 - \tau _ 1|} {8 \pi c _ 1 \m R ^ 2}~,
\ee
we get:
\bea
\label{Flargedist}
\begin{split}
F _ {q, -q} ^ {\d _ 1 \d _ 2} &= C_1 C _2 \m ^ {\d _ 1 + \d _ 2} e ^ {-\m q( \tau _ 2 -  \tau _ 1)} e ^ {- \D _ Q (\tau _ {out} - \tau _ {in})/R} 
\\
&\times
\bigg[1- q ^ 2\f {|\tau _ 2 - \tau _ 1|} {8 \pi c _ 1 \m R ^ 2} + \f {q(\delta_1+\delta_2)} {8 \pi c _ 1 \m ^ 2 R ^ 2} + \frac{3}{8\pi c_1\mu R}\(q^2+\frac{q(\delta_1+\delta_2)}{\mu R}\){\bf n_1}\cdot{\bf n_2}~ e^{-(\tau_2-\tau_1)/R}\bigg]~.  \nn
\end{split}
\eea
Once mapped onto the plane, the above expression gives rise to the following four-point function 
\bea
\label{4ptcylpl}
\begin{split}
\la \mathcal{O}_{-Q}(x_{out})\mathcal{O}_{-q,\delta_2}(x_2)\mathcal{O}_{q,\delta_1}(x_1)\mathcal{O}_Q(x_{in}) \ra  =
\l (\f{ Q}{ 2 \pi c_1 } + \frac{q}{4 \pi c_1 Q} \r ) ^ {\f{\d _ 1 + \d _ 2}{2}} \f{C_1 C _2 } {| x _ {out} |^{2 \D _ Q} \, | x _ {1} |^{\d _ 1} \, | x _ {2} |^{\d _ 2}}
~~~~ \\  \times ~
\l ( \f {|x _ 1|} {| x _ 2 |} \r ) ^ { q \sqrt{Q/2 \pi c _ 1} +q ^ 2/ \sqrt{32 \pi c _ 1 Q}}  
\bigg[ 1+\frac{3}{4}\f{1}{\sqrt{2\pi c_1 Q}}\(q^2+{q(\delta_1+\delta_2)}\sqrt{\f{2\pi c_1}{Q}}\)\f{{x_1}\cdot{x_2}}{x_2^2} \bigg]~,~~~
\end{split}
\eea
where we have kept terms at most of order $1/Q$.

It is instructive to derive the same result using the OPE in the $(x_{out} ~x_2)~(x_1~x_{in})$ channel.
To this end, consider first a general OPE for two scalar primaries   
\bea
\label{OPE}
\begin{split}
\mathcal{O}_2(x_2) \mathcal{O}_1(x_1)|0\ra &=& \sum_N\frac{\lambda_{12N}}{x_{21}^{\Delta_1+\Delta_2-\Delta_N}}\(1+\frac{\Delta_2+\Delta_N-\Delta_1}{2\Delta_N} x^\mu_{21}\p^1_\mu+\dots\) \mathcal{O}_N(x_1)|0\ra ~~~~~~\\
&=&  \sum_K\frac{\lambda_{12N}}{x_{21}^{\Delta_1+\Delta_2-\Delta_N}}\(1+i \frac{\Delta_2+\Delta_N-\Delta_1}{2\Delta_N} x^\mu_{21}P_\mu+\dots\) \mathcal{O}_N(x_1)|0\ra,~~~~~~
\end{split}
\eea
where the sum runs over primaries $\mathcal{O}_N$ with the appropriate internal quantum numbers. The contributions from descendants are fixed by the conformal symmetry, as indicated by the (scaling dimension-dependent) coefficients of terms, linear in the momentum operator \cite{Pappadopulo:2012jk}. (By the ellipses we denote the contributions from all other descendants, obtained by applying the momentum operator $n$ times, where $n\geq 2$).

Recalling that the Minkowskian Hermitian conjugation maps into inversion in radial quantization, we have
\be
\label{conjugation}
\(\mathcal{O}_4(x_4) \mathcal{O}_3(x_3)|0\ra\)^\dagger = \la 0 |\mathcal{O}_3(Ix_3)\mathcal{O}_4(I x_4)|x_3|^{-2\Delta_3}|x_4|^{-2\Delta_4}~,
\ee
where $I$ is the inversion operator, $I x^\mu \equiv x^\mu/x^2$.
Furthermore,  $P_\mu^\dagger =I P_\mu I=K_\mu$, so that the conjugate of eq.~\eqref{OPE} is
\bea
\label{ope2}
\begin{split}
\la 0 |\mathcal{O}_3(I x_3)\mathcal{O}_4(I x_4) &=  \sum_N\lambda_{34N} \frac{|x_3|^{2\Delta_3}|x_4|^{2\Delta_4}}{|x_3|^{2\Delta_N}x_{43}^{\Delta_3+\Delta_4-\Delta_N}} \\ &\times \la 0 |\mathcal{O}_N(Ix_3)\(1-i \frac{\Delta_4+\Delta_N-\Delta_3}{2\Delta_N} x^\mu_{43}K_\mu+\dots\)~.
\end{split}
\eea
For the four-point function (\ref{4ptplane}) the OPE in the $(x_{out} ~x_2)~(x_1~x_{in})$ channel corresponds to inserting intermediate states with charge $Q + q$. For all these intermediate states, 
eq.~(\ref{F}) clearly implies that the fusion  coefficients  are controlled by two coefficients $C_{1,2}$. The leading contribution comes from the exchange of the lowest dimension scalar primary of charge $Q+q$, whose 3-point function we discussed in the previous subsection.
The correlator (\ref{4ptplane}) can be readily evaluated using eqs.~\eqref{OPE} and \eqref{ope2} with the identification: $x_1=x_{in},~ x_2= x_1,~x_3=I x_{out},~x_4=I x_2$ as well as $\Delta_1 = \Delta_Q,~\Delta_2=\delta_1,~\Delta_3=\Delta_Q,~\Delta_4=\delta_2$ and $\Delta_N = \Delta_{Q+q}$.\footnote{The subleading terms will correspond to the combined effect of some of its descendants and the excited states of spin $\ell$.} In the limit $x_{out}\to \infty,~x_{in}\to 0$, the result becomes\footnote{To arrive at \eqref{4ptpl}, we use $[K_\mu, P_\nu] = -2 i \big(g_{\mu\nu}D+J_{\mu\nu}\big)$ and $J_{\mu\nu}\mathcal{O}_{Q+q}(0)|0\ra=0$, that follows from the fact that $\mathcal{O}_{Q+q}$ is a scalar.}
\bea
\label{4ptpl}
\begin{split}
\la 0 |\mathcal{O}_{-Q}(x_{out})\mathcal{O}_{-q,\delta_2}(x_2)\mathcal{O}_{q,\delta_1}(x_1)\mathcal{O}_Q(x_{in})  |0\ra=  \(\f {|x _ 1|} {| x _ 2 |}\)^{\Delta_{Q+q}-\Delta_Q} \f{\lambda_{1QQ+q}~\lambda_{2QQ+q} } {| x _ {out} |^{2 \D _ Q} \, | x _ {1} |^{\d _ 1} \, | x _ {2} |^{\d _ 2}}~~~~~\\  \\ \times \(1+\frac{\(\delta_2+\Delta_{Q+q}-\Delta_{Q}\) \(\delta_1+\Delta_{Q+q}-\Delta_{Q}\)}{2\Delta_{Q+q}}~ \frac{x_2\cdot x_1}{x_2^2}\).~~~~~~~~~~~~~~~~
\end{split}
\eea
One can straightforwardly show, using the explicit (leading-order) expression for $\Delta(Q)$ in eq.~\eqref{eqdelta0}, that the result \eqref{4ptpl} obtained with the help of the OPE exactly agrees with eq.~\eqref{4ptcylpl} obtained from the direct calculation. In particular,
\be
\D _ {Q+q} - \D _ Q \approx q \f {\p \D _ Q} {\p Q} + \f {q ^ 2} {2} \, \f {\p^2 \D _ Q} {\p Q^2} = q \sqrt{\f{Q}{2 \pi c _ 1}} + 
\f{q ^ 2} {\sqrt{32\pi c _ 1 Q}},
\ee
precisely matches with the power of $|x _ 1|/|x _ 2|$ in (\ref{4ptcylpl}).

\subsubsection*{Small separations}

When the distance between the two insertions is much smaller than the radius $R$ of the cylinder, the propagator of the Goldstone mode, $\la \pi (\tau_2, {\bf n}_2) \pi (\tau_1,{\bf n}_1) \ra$, can be approximated by its flat-space expression. In that case, the desired correlator becomes
\bea
F _ {q, -q } ^ {\d _ 1 \d _ 2} &=&
C_1 C _2 (- i \m) ^ {\d _ 1 + \d _ 2} e ^ {\m (q _ 1 \tau _ 1 + q _ 2 \tau _ 2)} e ^ {- \D _ Q (\tau _ {out} - \tau _ {in})/R} \nn \\ &\times& \bigg ( 1 +  \f {1} {\sqrt{2\pi c _ 1 Q}} \, 
\f {q^2+{q(\delta_1+\delta_2)}\sqrt{\f{2\pi c_1}{Q}}} {\sqrt{(\tau _ {2} - \tau _ {1})^2 / R ^ 2+\t_{12} ^ 2}} \bigg ), 
\label{4pt_sphere_ex}
\eea
with $\t _{12} \ll 1$. 

Mapping \eqref{4pt_sphere_ex} onto the plane results in the following expression for the four-point function
\bea
\label{4ptshort}
\begin{split}
\la {\cal O}_{- Q}(x_{out}){\cal O}_{-q, \d_2}(x_2){\cal O}_{q, \d_1}(x_1){\cal O}_{Q}(x_{in}) \ra \underset{
\ba{lll}
\scriptstyle x_{out} \to \infty\\
\scriptstyle x_{in} \to 0
\ea
}
{=} 
\l (\f{ Q}{ 2 \pi c_1 } \r ) ^ {\f{\d _ 1 + \d _ 2}{2}} \l ( \f {|x _ 1|} {| x _ 2 |} \r ) ^ {q \sqrt{Q/2 \pi c _ 1}} \\ \times~
\f{C_1 C _2 } {| x _ {out} |^{2 \D _ Q} \, | x _ {1} |^{\d _ 1} \, | x _ {2} |^{\d _ 2}} 
\l ( 1 +  \f {1 } {\sqrt{ 2 \pi c_1 Q}} \, 
\f {q ^ 2} {\sqrt{\log ^ 2 (| x _ 2 | / | x_1 |)+4(1 - \mathbf{n}_1\cdot\mathbf{n}_2)}} \r ). 
\end{split}
\eea
One can straightforwardly check that the above expression is consistent with the ($x _ {out}\to \infty$, $x_ {in} \to 0$ limit of) the general form of a CFT four-point function
\be
\la {\cal O}_{- Q}(x_{out}){\cal O}_{-q, \d_2}(x_2){\cal O}_{q, \d_1}(x_1){\cal O}_{Q}(x_{in}) \ra = \f {f (u,v)} 
{| x _ {out,in} | ^ {2 \D - \d _ 1 - \d _ 2} | x _ {out,1}| ^ {\d _1} | x _ {out,2} | ^ {\d _2} 
| x _ {in,1} | ^ {\d _1} | x _ {in,2} | ^ {\d _2}}~. \nn
\ee
Here the two conformal ratios $u$ and $v$ have been defined as
\be
u = \f {x^2_{1,2} x^2 _ {out,in}} {x_{out,1}^2 x _ {2,in} ^ 2} \to \f {|x _ 1|^2} {|x _ 2|^2}, ~~ 
v = \f {x^2_{1,in} x^2 _ {out,2}} {x_{out,1}^2 x _ {2,in} ^ 2} \to 1 + u - 2 \sqrt{u} \, \mathbf{n}_1\cdot\mathbf{n}_2,
\ee
and the function $f (u,v)$ is given by
\be
f (u,v) = \l (\f{ 3 Q}{ \pi c_1 } \r ) ^ {\f{\d _ 1 + \d _ 2}{2}} C _ 1 C _ 2 \, u ^ {q \sqrt{Q/8 \pi c _ 1}}
\l ( 1 + \sqrt{ \f {2 } { \pi c_1 Q}} \, 
\f {q ^ 2} {\sqrt{\log ^ 2 u + 16\l ( 1- \f {1+u-v} {2 \sqrt{u}}\r ) }} \r ).
\ee

As expected, the expansion in (\ref{4pt_sphere_ex}) is only consistent provided  the following relation holds:
\be
\sqrt{(\tau _ {2} - \tau _ {1})^2 / R ^ 2 + \t ^ 2} \gg   \f {q ^ 2} {\sqrt{2 \pi c _1 Q}}~.
\ee
This agrees with our expectation that new (gapped) degrees of freedom come in around distance scales of order $\m^{-1} \sim  R/Q^{1/2}$. Notice as well, that the singularity structure of the correlator \eqref{4ptshort} suggests that it does not correspond to the OPE, applicable in the limit $x_2\to x_1$. We thus expect that the latter only becomes convergent for even shorter distances $|x_2-x_1|\ll \mu^{-1}$, where the EFT at hand is not applicable.

Our final comment concerns the procedure of operator matching for the case of non-Abelian symmetries. We illustrate it on the example of the $SO(3)$ group, discussed in the previous section. We will focus on scalar operators, so that each representation $\kappa _ \ell$ of $SO(3)$ (labelled by an integer $\ell$) is decomposed into $2 \ell + 1$ trivial representations of the unbroken group. Therefore, an operator characterized by the quantum number $\ell$ and the dimension $\d$ is represented by (see eq.~\eqref{reps2})
\be
\mc O _ {\ell, \d} = (\c _ a \c ^ a) ^ {\d/2} \kappa _ \ell (e ^ {i \c \wh Q _ 3} e ^ {i \pi ^ I \wh Q _ I})\tilde \phi,
\ee
with $\c _ a$ defined in (\ref{chi_pi_def}) and $\tilde \phi$ being a constant $(2 \ell + 1)$ component vector. The vector $\tilde \phi$ is thus characterized by $2\ell +1$ input constants $C_m$ ($m=-\ell, \dots,+\ell$), which are the $SO(3)$ analogues of the constant $C$ appearing in eq.~(\ref{IRop}) for the abelian case. These $2\ell+1$ constants are precisely associated to the $2\ell+1$ irreducible representations that arise in the tensor product ${\mathbf \ell}\otimes {\bf Q}_3$ which arise when considering the action of $ O _ {\ell, \d}$ on our ground state $|Q_3\rangle$, which corresponds to a
${\bf Q}_3$ irreducible representation of $SO(3)$ (notice that in our approach $Q_3\gg \ell$ by construction).

\section{Conclusions \label{sec_conclusions}}

Conformal field theories  simplify substantially in the limit of large quantum numbers, many of their features becoming amenable to semiclassical analysis. In particular, focussing on 3-dimensional CFT with a global $U(1)$ symmetry, ref.~\cite{Hellerman:2015nra} has recently shown how the properties of the lowest-dimension large charge operators can be studied, through radial quantization, by considering the system on a spatial two-sphere of radius $R$ in a state with large $U(1)$ charge.
There is a sense in which systems at non-zero charge density can be treated as those where, in addition to the Lorentz boosts, both the internal group and time translations are spontaneously broken, while a certain linear combination of the two remains linearly realized. This corresponds to the symmetry breaking pattern of a generalized (conformal) superfluid, discussed in Sec.~\ref{general_strategy}.
Below the energy set by the charge density, the CFT of interest is described by the Goldstone boson that nonlinearly realizes the broken symmetries and becomes more and more weakly coupled as the charge is increased. The dynamics of that Goldstone is determined by the symmetry breaking pattern at hand and is largely independent of the precise details of the UV CFT, leading to a significant amount of universality in the predictions. 

This paper serves to present the above picture from a systematic perspective, which allows for a straightforward extension to more involved cases of CFTs with multiple (Abelian or non-Abelian) internal charges. We have provided arguments that show how the fixed-charge path integral formally imposes the description in terms of spontaneously  broken symmetries and the associated Goldstone bosons. This is exemplified by the semiclassical derivation of the (well-known) spectrum of the rigid rotor at large angular momentum, which contains many of the conceptual ingredients of the more involved case of CFTs at large internal charge. We have provided a systematic derivation of the effective action for a $d$-dimensional conformal superfluid by employing the CCWZ methodology for spontaneously broken space-time and internal symmetries. While not particularly beneficial for the simplest case with $U(1)$ symmetry, the CCWZ construction becomes crucial once more involved non-linearly realized internal and/or spacetime symmetries are considered. As an example, we have generalized the results of ref.~\cite{Hellerman:2015nra} to the cases of CFTs with $U(1)\times U(1)$ and $SO(3)$ global symmetries.

Furthermore, we have provided an opening discussion of how extra CFT data, i.e  $n$-point functions with  two insertions of large charge operators and any insertion of operators with finite  dimensions and finite  charge, can be accessed with the help of the semiclassical analysis. We have shown, focussing on the simplest example of a $U(1)$ symmetry group, that there exists a universal scaling of the fusion coefficients with the large charge $Q$. In particular, we have provided a prescription for calculating, semiclassically, four-point functions of the form $\la {\cal O}_{- Q}(\infty){\cal O}_{-q, \d_2}(x_2){\cal O}_{q , \d_1}(x_1){\cal O}_{Q}(0)\ra$, with $q\ll Q$. For sufficiently large separations on the cylinder, $|x_2-x_1|\gg R$, we have  cross-checked that the semiclassical result exactly reproduces the one obtained by  applying conformal invariance and the  operator product expansion.~In the OPE calculation, the leading contribution to the four-point function  comes from the exchange of a scalar primary of charge $Q+q$ and scaling dimension $\Delta(Q+q)$. It should be stressed that all of the subleading effects due to descendants and  higher-spin primaries are remarkably combined, at the leading order in the large charge expansion, into the dependence on just two unknown constants $C_1$ and $C_2$ defined in eq.~\eqref{IRop}. In contrast, for small separations $|x_2-x_1|\lsim R$, our result (although consistent with conformal invariance) does not reproduce the corresponding short distance   OPE. This suggests that the latter is only convergent for separations on the cylinder of order $1/\m$ or less, which lie outside the range of applicability of our effective field theory. We plan to study the various limits of higher point correlators and their expansion in the various intermediate states in more detail in a further publication.

Our work can be extended in several directions. First and foremost, it would be interesting to understand to what extent the CCWZ construction outlined in Sec.~\ref{sec_coset} can be useful to explore the sectors of CFT with large \textit{spin}, as opposed to internal charge. Another important problem is pushing forward our preliminary discussion of Sec.~\ref{npt} on the semiclassical calculation of the various $n$-point functions of scalar primaries and/or conserved currents. Last but not least, it would be nice generalize our results to non-relativistic CFTs at large global charge, with an eye towards applying the methodology developed in this work to condensed matter systems.

\subsection*{Acknowledgements}
We acknowledge helpful discussions with A. de la Fuente, A. Nicolis, J. Penedones, F. Piazza,  S. Rychkov and M. Serone. The work of A.M. is supported by the Swiss National Science Foundation grant Ambizione. D.P. is partially supported
by the Swiss National Science Foundation under grant 200020-150060.
The work of R.R. is partially supported by the Swiss National Science foundation under grant 200020-150060 and
also through the NCCR SwissMAP. The work of F.S. is partially supported by grant no. 615203 from the European Research Council under the FP7.

\appendices

\section{The $R \to \infty$ limit of the cylinder \label{R_infty_limit}}
\label{appB}

We start out with a particular parametrization of the cylinder in which the metric is conformally flat
\be
d s _ {\rm cyl} ^ 2 = R ^2~ \f {dx _ 0 ^ 2 + dx _ i ^ 2} {x _ 0 ^ 2 + x _ i ^ 2}, ~~~ i = 1,\dots, d-1~.
\ee
In these coordinates, the generators of the conformal group have the standard differential representation 
\bea
P _ \m & = & - i \p _ \m, \nn \\
D & = & i x ^ \m \p _ \m, \nn \\
J _ { \m \n } & = & i \l ( x _ \m \p _ \n - x _ \n \p _ \m \r ), \nn \\
K _ \m & = & - i \l ( 2 x _ \m x ^ \n \p _ \n -x ^ 2 \p _ \m \r ).
\label{conformal_diff_operators}
\eea
We use the following convention for commutation relations
\bea
\begin{split}
\l [ D, P _ \m \r ] & =  - i P _ \m,  \\
\l [ J _ {\lambda \s}, P _ \m \r ] & =  i g _ {\s \m} P _ \lambda - i g _ {\lambda \m} P _ \s,  \\
\l [ K _ \m, P _ \n \r ] & =  - 2 i \l ( g _ {\m \n} D + J _ {\m \n} \r ),  \\
\l [ D, K _ \m \r ] & =  i K _ \m,  \\
\l [ J _ {\m \n}, M _ {\lambda \s} \r ] & =  i \l ( J _ {\m \s} g _ {\n \lambda} + J _ {\n \lambda} g _ {\m \s} -
J _ {\n \s} g _ {\m \lambda} - J _ {\m \lambda} g _ {\n \s} \r ),  \\
\l [ J _ {\lambda \s}, K _ \m \r ] & =  i g _ {\s \m} K _ \lambda - i g _ {\lambda \m} K _ \s~.
\label{conf_cr}
\end{split}
\eea
Consider now a set of coordinates $(\tau, y _ i)$, defined through the following relations
\be
x _ 0 = \sqrt{R^2 e ^ {2\tau/R} - y _ i ^2}, ~~~ x _ i = y _ i~.
\ee
In these new coordinates, the metric becomes flat in the $R \to \infty$ limit
\be
d s _ {\text cyl} ^ 2 =d\tau ^ 2 + dy _ i ^ 2 + \mc O (1/R)~,
\ee
and it is a straightforward exercise to show that the corresponding conformal generators (which we denote by symbols with hats) can be obtained as a $R \to \infty$ contraction of the algebra (\ref{conformal_diff_operators}). For instance,
\be
J _ {0i} = i \l (x _ 0 \f{\p}{\p x _ i} - x _ i \f{\p}{\p x _ 0} \r ) = i \sqrt{R ^ 2 e ^ {2 \tau / R} - y _ j ^ 2} \, \f {\p} {\p y _ i} \underset{R \to \infty}{=}
- R \widehat P _ i~.
\ee
Continuing in the same manner, the map between the two sets of generators can be found to be given by the following expressions
\bea
D & = & - R \wh P _ 0, ~~ J _ {i j} = \wh J _ {i j}, ~~ J _ {0 i} = - R \wh P _ i, \\
P _ 0 & = & \wh P _ 0 + \f {\wh D} {R} + \f {\wh K _ 0} {2 R ^2}, ~~ K _ 0 = \f{1} {2} \wh K _ 0 - R {\wh D} + {R ^2} {\wh P _ 0}, \\
P _ i & = & \wh P _ i + \f {\wh J _ {0i}} {R} - \f {\wh K _ i} {2 R ^2}, ~~  K _ i = \f{1} {2} \wh K _ i + R {\wh J _ {0i}} - {R ^2} {\wh P _ i}~.
\eea

\section{Casimir energies on a sphere}
\label{app}

In this Appendix we further elaborate on the results of Secs. \ref{u1} and \ref{u1timesu1} for the one-loop corrections to the scaling dimensions of the lowest-lying operators. 

Consider an operator $\hat O$, defined on a smooth background manifold. For the purposes of regularizing its determinant, it is convenient to consider the following quantity
\be
\zeta(s|\hat O)= \text{Tr} \, \hat O^{-s}.
\label{eq:def_zeta}
\ee
This generalizes the standard Riemann zeta function $\zeta(s)$, which corresponds to the special case of an operator whose set of eigenvalues coincides with $\mathbb{Z}^+$. 

For a generic $\hat O$, the sum in \eqref{eq:def_zeta} is convergent for a sufficiently large real part of $s$; after evaluating it in the domain of convergence, one can analytically continue to $s=0$.
The determinant of $ \hat O$ can then be found through the following identity
\be
\ln \det  \hat O =  -\l.\frac{d}{ds}\zeta\l(s |\hat O\r)\r|_{s=0} ~.
\label{eq:lndetFU1}
\ee
In what follows we will apply this procedure to compute Casimir energies for the systems of Secs. \ref{u1} and \ref{u1timesu1}.

\subsection{\texorpdfstring{$U(1)$ } { }}
In the case of a CFT with a global $U(1)$ group, the operator of interest reads
\be
\hat F  = -\f{1}{\Lambda^2} \l(\p_\tau^2 + \f{1}{2} \Delta_{S^2}  \r),
\label{eq:GammaU1}
\ee
and a potential difficulty arises due to the presence of the zero mode with $\p_\tau=0$ and $l=0$.\footnote{We choose to normalize $\hat F$ by an arbitrary scale $\Lambda$ in order to make its eigenvalues dimensionless. The additive constant does not affect our results, since we are only interested in the large-$T$ behavior of the amplitude.} In order to deal with this IR divergence, we will formally regulate the spectrum by introducing a small mass term in \eqref{eq:GammaU1}. Such a regulator clearly breaks conformal invariance, but we will see that it will eventually fall out from the calculation.

The calculation therefore amounts to evaluating $\zeta(s|\hat F)$ for an arbitrary $s$ in the domain of convergence and then analytically continuing to the point of interest $s=0$. For a finite temporal interval, $-T/2 <t<T/2$, we have 
\bea 
\label{eqzetafunctu1}
\zeta(s|\hat F) & = & T \int \f{\text{d}\omega}{2 \pi} \sum_{l=0}^\infty (2l+1) \l[\f{\omega^2}{\Lambda^2} + \f{1}{2} \f{l(l+1)}{(\Lambda R)^2} + \f{m^2}{\Lambda^2} \r]^{-s} \label{eq:zetaF}\nn \\
&=&  \f{T}{R} (\Lambda R)^{2s} \f{\Gamma(s-1/2)}{\Gamma(s)} \f{2^{s-1/2}}{2 \sqrt{\pi}}  \sum_{l=0}^\infty (2l+1) \l[l(l+1)+ 2(mR)^2\r]^{1/2-s} \quad \nn \\
&\equiv&  \f{T}{R} \f{F(s)}{\Gamma(s)}~.
\eea
The way $F(s)$ has been defined makes it regular at $s=0$,\footnote{This we haven't shown yet, but it will become clear \textit{a posteriori}, once we analytically continue the sum in \eqref{eqzetafunctu1} to $s=0$.} and using \eqref{eq:lndetFU1} the desired determinant reads\footnote{Here we have used
\be
\l. \f{1}{\Gamma(s)}\r|_{s=0}=0 \quad \text{and} \quad \l.\f{\text{d}}{\text{d}s}\l( \f{1}{\Gamma(s)}\r)\r|_{s=0}=1 ~.
\ee }
\be
\ln \det \hat F = -  \f{T}{R}  \l. F(s)\r|_{s=0}.
\ee
Now, at the point of interest $s=0$, the $l=0$ piece of the infinite sum in the definition of $F(s)$ is manifestly regular for $m\to 0$. This allows to remove the IR regulator, yielding a simplified expression
\be
\ln \det \hat F = - \f{T}{R} \l.(\Lambda R)^{2s} \Gamma\(s-\f{1}{2}\)\f{2^{s-1/2}}{2 \sqrt{\pi}}  \zeta\(s-\f{1}{2} \bigg |-\Delta_{S^2}\)\r|_{s=0}~.
\label{eq:zetaFzetaLU1}
\ee
The task of computing the regularized determinant thus reduces to evaluating $\zeta(s |-\Delta_{S^2})$ for an arbitrary $s$, and then analytically continuing to $ s=-1/2$. To this end, it is useful to define an auxiliary function
\be
f(s;a,b,c) = \sum_{l=1}^\infty l^{-s+b} (l+a)^{-s+c}
\label{eq:auxfunction}
\ee 
in terms of which 
\be
\label{eq:zetalaplacian}
\zeta(s|-\Delta_{S^2})=\sum_{l=1}^\infty \f{2l+1}{\l[l(l+1)\r]^s} = f(s,1,0,1)+f(s,1,1,0)~.
\ee
The latter expression can be conveniently rewritten as follows. First, one splits the sum in \eqref{eq:auxfunction} as
\be
\label{eq:auxfunction1}
f(s;a,b,c) = \sum_{l=1}^{[a]} l^{-2s+b+c} \l(1+\f{a}{l}\r)^{-s+c} + \sum_{l=[a]+1}^{\infty} l^{-2s+b+c} \l(1+\f{a}{l}\r)^{-s+c} ~,
\ee    
where $[a]$ denotes the integer part of $a$.
Applying the binomial expansion to the second term yields
\be \begin{aligned}
f(s;a,b,c) \,=&{\color{white}+}\, \sum_{l=1}^{[a]} l^{-2s+b+c} \l(1+\f{a}{l}\r)^{-s+c} \\ 
&+\, \sum_{k=0}^{\infty} \f{\Gamma(1-s+c)}{k!\,\Gamma(1-s-k+c)} a^k  \l[\zeta(2s+k-b-c) - \sum_{l=1}^{[a]} l^{-2s-k+b+c}\r]~,
\end{aligned} \ee
where we have extended the sum in $k$ from $0\leq k\leq -s+c$ to $0\leq k\leq \infty$. This does not change the result as $\Gamma(1-s-k+c)^{-1}$ vanishes for $k >-s+c$~.
Our case of interest corresponds to $[a]=1$ in which case \eqref{eq:zetalaplacian} reduces to
\be \begin{aligned}
\zeta\l(s|-\Delta_{S^2}\r) = 2^{-s+1} + 2^{-s} &+ \sum_{k=0}^\infty \f{\Gamma(1-s)}{k! \, \Gamma(1-s-k)} \l[ \zeta(2s+k-1)-1\r] \\
&+ \sum_{k=0}^\infty \f{\Gamma(2-s)}{k! \, \Gamma(2-s-k)} \l[ \zeta(2s+k-1)-1\r]~. \end{aligned}
\label{eq:zeta1}
\ee
For the point of interest, $s= -1/2$, the only apparent divergence in the above sum occurs for $k=3$ corresponding to the pole of the zeta function ($\zeta(2 s+k-1)\sim (2 s+k-2)^{-1}$ for $s\to -1/2$ and $k\to 3$.) However, upon closer inspection of this term, one can see that its prefactor is itself proportional to $(2 s+k-2)$, removing the would-be divergence. The expression in Eq. \eqref{eq:zeta1} then numerically evaluates to\footnote{This result is numerically off from the analogous calculation of Ref. \cite{Hellerman:2015nra}. The reason is that upon evaluating the functional determinant, the authors of \cite{Hellerman:2015nra} use the zeta function regularization in a non-covariant way. In particular, when evaluating $\zeta(-1/2|-\Delta_{S^2})=\sum_{l=1}^\infty (2l+1) \sqrt{l(l+1)}$, the sum is split \textit{before} performing the regularization. This can be a tricky step, given that the separate contributions do not converge. The following example illustrates how this procedure can go wrong (for more see~\cite{Bilal:2013iva,Assel:2015nca,Monin:2016bwf}):
\bea
\sum_{l=1}^\infty (l+1) &\stackrel{\text{$\zeta$ regularization}}{\rightarrow}& \zeta(-1)-1 = -\f{1}{12} - 1 \\
\sum_{l=1}^\infty l + \sum_{l=1}^\infty 1  &\stackrel{\text{$\zeta$ regularization}}{\rightarrow}& \zeta(-1)+\zeta(0)=-\f{1}{12}-\f{1}{2} ~.
\eea
}
\be
\zeta(-1/2|-\Delta_{S^2}) = -0.265096~,
\ee
which directly leads to the result in Eq. \eqref{eqdelta0}. 

{
 The function $\zeta(-1/2|-\Delta_{S^2}) $ can also be found using the results of ref.~\cite{Monin:2016bwf}. There it was shown the zeta function can be found as $t$-independent term in the asymptotic 
$t \to 0$ expansion of the regularized sum, leading to
\bea
\sum _ {l=0} ^ {\infty} (2 l + 1) \sqrt{l (l + 1)} \, e ^ {- t \, l (l + 1)}  &\underset{t \to 0}{=} &
\sum _ {l=0} ^ {\infty} \l (2 l ^ 2 + 2 l + \f {1}{4} \r ) e ^ {- t \, l (l + 1)}  \nn \\&-&
\sum _ {l=0} ^ {\infty} (2 l + 1) \sqrt{l (l + 1)} - 2l ^ 2 -2 l - \f{1} {4}~.
\eea
The second term on the r.h.s. is convergent and can be computed numerically, while the asymptotic expansion of the first one is reproduced using the Euler-Maclaurin formula. As a result, 
\be
\sum _ {l=0} ^ {\infty} (2 l + 1) \sqrt{l (l + 1)} \, e ^ {- t \, l (l + 1)}  \underset{t \to 0}{=} 
\f {\sqrt{\pi}} {2 t ^ {3/2}} - \f {1} {4} - 0.015096~.
\ee

}


\subsection{\texorpdfstring{$U(1)\times U(1)$ } { }}

For CFTs with a $U(1)\times U(1)$ internal group, the functional determinant we need to evaluate is given in \eqref{eq:GammaU1xU1}. The analog of Eq. \eqref{eqzetafunctu1} in zeta function regularization then reads
\be
\label{eqzetaG}
\zeta(s|\hat G)=T  \int \f{ \text{d} k}{2 \pi} \sum_{l=0}^\infty (2l+1)\l[ \l(\f{k^2}{\Lambda^2} + \frac{1}{2}\f{l(l+1)}{(R \Lambda)^2} \r) \l(\f{k^2}{\Lambda^2} + c^2_-\f{l(l+1)}{(R \Lambda)^2} \r)\r]^{-s},
\ee 
where $\hat G =(-\p_\tau^2-\Delta_{S^2}/2) (-\p_\tau^2- c^2_-\Delta_{S^2})/\Lambda^4$. The expression under the integral is conveniently manipulated using Feynman parametrization:
\be
 \l(\f{k^2}{\Lambda^2} + \frac{1}{2}\f{l(l+1)}{(R \Lambda)^2} \r)^{-s} \l(\f{k^2}{\Lambda^2} + c^2_-\f{l(l+1)}{(R \Lambda)^2}\r)^{-s} =\frac{\Gamma(2s)}{\Gamma(s)^2}\int_0^1 dx \frac{x^{s-1}(1-x)^{s-1}}{C^{2s}}~,\nn
\ee
where
\be
C\equiv \frac{1}{\Lambda^2}\l[k^2+\frac{l(l+1)}{2R^2}\l[x+2 c^2_-(1-x)\r]\r]~.
\ee
It proves useful to further rescale the integration variable as $k\to \l[x+2 c^2_-(1-x)\r]^{1/2}k$, which puts the expression in Eq. \eqref{eqzetaG} into the following form
\be
\zeta(s|\hat G)=\zeta(2s|\hat F)\frac{\Gamma(2s)}{\Gamma(s)^2}\int_0^1dx \frac{x^{s-1}(1-x)^{s-1}}{\l[x+2 c^2_-(1-x)\r]^{2s-1/2}}
\ee
The integral over the Feynman parameter $x$ can be expressed in terms of the hypergeometric function, leading to the final result 
\bea
\zeta(s|\hat G)&=&  \zeta(2s|\hat F)~ \,(2 c^2_-)^{1/2-2s} ~ \,_2 F_1\l[s,2s-\f{1}{2};2s;1-\f{1}{2 c^2_-}\r]~.
\eea
We are interested in the $s\to 0$ limit of the expression on the rhs. To extract the behavior of  $\zeta(2s|\hat F)$ in this limit, we note that it can be expressed through the following infinite series
\be
_2 F_1\l[s,2s-\f{1}{2};2s;1-\f{1}{2 c^2_-}\r]=1+\sum_{k=1}^\infty \f{\Gamma(s+k) \Gamma(2s-1/2+k) \Gamma(2s)}{k! \,\Gamma(s) \Gamma(2s-1/2) \Gamma(2s+k)} \l(1-\f{1}{2 c^2_-}\r)^k~,
\ee
which is regular in the limit $s\to 0$:
\bea
\l. _2 F_1\l[s,2s-\f{1}{2};2s;1-\f{1}{2 c^2_-}\r] \r|_{s=0}&=&1+\f{1}{2} \,\sum_{k=1}^\infty \f{ \Gamma(k-1/2)}{k! \,\Gamma(-1/2)} \l(1-\f{1}{2 c^2_-}\r)^k \\
&=& \f{1}{2}\l(1+\f{1}{\sqrt{2 c^2_-}}\r)~.
\eea
Recalling that $\zeta(2s|\hat F)|_{s\to 0}=0$ and $\zeta'(2s|\hat F)|_{s\to 0}=- 2 \ln \det \hat F$ (where $'=d/ds$), we arrive at the final result
\bea
\ln \det \hat G &=& -\l. \zeta'(s| \hat G)\r|_{s=0} =  \l(1+\sqrt{2} c^2_-\r) \ln \det \hat F~.
\eea
This leads to the expression in eq. \eqref{eqdelta0u1xu1}.

\renewcommand{\em}{}
\bibliographystyle{utphys}
\addcontentsline{toc}{section}{References}
\bibliography{large_charge}

\providecommand{\href}[2]{#2}\begingroup\raggedright\begin{thebibliography}{10}

\bibitem{Banks:1981nn}
T.~Banks and A.~Zaks, ``{On the Phase Structure of Vector-Like Gauge Theories
  with Massless Fermions},''
\href{http://dx.doi.org/10.1016/0550-3213(82)90035-9}{{\em Nucl. Phys.} {\bf
  B196} (1982)  189--204}.

\bibitem{Wilson:1972cf}
K.~G. Wilson, ``{Quantum field theory models in less than four-dimensions},''
\href{http://dx.doi.org/10.1103/PhysRevD.7.2911}{{\em Phys. Rev.} {\bf D7}
  (1973)  2911--2926}.

\bibitem{Seiberg:1994pq}
N.~Seiberg, ``{Electric - magnetic duality in supersymmetric nonAbelian gauge
  theories},'' \href{http://dx.doi.org/10.1016/0550-3213(94)00023-8}{{\em Nucl.
  Phys.} {\bf B435} (1995)  129--146},
\href{http://arxiv.org/abs/hep-th/9411149}{{\tt arXiv:hep-th/9411149
  [hep-th]}}.

\bibitem{Maldacena:1997re}
J.~M. Maldacena, ``{The Large N limit of superconformal field theories and
  supergravity},'' \href{http://dx.doi.org/10.1023/A:1026654312961}{{\em Int.
  J. Theor. Phys.} {\bf 38} (1999)  1113--1133},
  \href{http://arxiv.org/abs/hep-th/9711200}{{\tt arXiv:hep-th/9711200
  [hep-th]}}.
[Adv. Theor. Math. Phys.2,231(1998)].

\bibitem{Ferrara:1973yt}
S.~Ferrara, A.~F. Grillo, and R.~Gatto, ``{Tensor representations of conformal
  algebra and conformally covariant operator product expansion},''
\href{http://dx.doi.org/10.1016/0003-4916(73)90446-6}{{\em Annals Phys.} {\bf
  76} (1973)  161--188}.

\bibitem{Polyakov:1974gs}
A.~M. Polyakov, ``{Nonhamiltonian approach to conformal quantum field
  theory},''
{\em Zh. Eksp. Teor. Fiz.} {\bf 66} (1974)  23--42.

\bibitem{El-Showk:2014dwa}
S.~El-Showk, M.~F. Paulos, D.~Poland, S.~Rychkov, D.~Simmons-Duffin, and
  A.~Vichi, ``{Solving the 3d Ising Model with the Conformal Bootstrap II.
  c-Minimization and Precise Critical Exponents},''
  \href{http://dx.doi.org/10.1007/s10955-014-1042-7}{{\em J. Stat. Phys.} {\bf
  157} (2014)  869},
\href{http://arxiv.org/abs/1403.4545}{{\tt arXiv:1403.4545 [hep-th]}}.

\bibitem{Hellerman:2015nra}
S.~Hellerman, D.~Orlando, S.~Reffert, and M.~Watanabe, ``{On the CFT Operator
  Spectrum at Large Global Charge},''
  \href{http://dx.doi.org/10.1007/JHEP12(2015)071}{{\em JHEP} {\bf 12} (2015)
  071},
\href{http://arxiv.org/abs/1505.01537}{{\tt arXiv:1505.01537 [hep-th]}}.

\bibitem{Nicolis:2015sra}
A.~Nicolis, R.~Penco, F.~Piazza, and R.~Rattazzi, ``{Zoology of condensed
  matter: Framids, ordinary stuff, extra-ordinary stuff},''
  \href{http://dx.doi.org/10.1007/JHEP06(2015)155}{{\em JHEP} {\bf 06} (2015)
  155},
\href{http://arxiv.org/abs/1501.03845}{{\tt arXiv:1501.03845 [hep-th]}}.

\bibitem{Coleman:1969sm}
S.~R. Coleman, J.~Wess, and B.~Zumino, ``{Structure of phenomenological
  Lagrangians. 1.},''
\href{http://dx.doi.org/10.1103/PhysRev.177.2239}{{\em Phys. Rev.} {\bf 177}
  (1969)  2239--2247}.

\bibitem{Callan:1969sn}
C.~G. Callan, Jr., S.~R. Coleman, J.~Wess, and B.~Zumino, ``{Structure of
  phenomenological Lagrangians. 2.},''
\href{http://dx.doi.org/10.1103/PhysRev.177.2247}{{\em Phys. Rev.} {\bf 177}
  (1969)  2247--2250}.

\bibitem{Salam:1969rq}
A.~Salam and J.~A. Strathdee, ``{Nonlinear realizations. 1: The Role of
  Goldstone bosons},''
\href{http://dx.doi.org/10.1103/PhysRev.184.1750}{{\em Phys. Rev.} {\bf 184}
  (1969)  1750--1759}.

\bibitem{Nicolis:2012vf}
A.~Nicolis and F.~Piazza, ``{Implications of Relativity on Nonrelativistic
  Goldstone Theorems: Gapped Excitations at Finite Charge Density},''
  \href{http://dx.doi.org/10.1103/PhysRevLett.110.011602,
  10.1103/PhysRevLett.110.039901}{{\em Phys. Rev. Lett.} {\bf 110} (2013)
  no.~1, 011602}, \href{http://arxiv.org/abs/1204.1570}{{\tt arXiv:1204.1570
  [hep-th]}}.
[Addendum: Phys. Rev. Lett.110,039901(2013)].

\bibitem{Pappadopulo:2012jk}
D.~Pappadopulo, S.~Rychkov, J.~Espin, and R.~Rattazzi, ``{OPE Convergence in
  Conformal Field Theory},''
  \href{http://dx.doi.org/10.1103/PhysRevD.86.105043}{{\em Phys. Rev.} {\bf
  D86} (2012)  105043},
\href{http://arxiv.org/abs/1208.6449}{{\tt arXiv:1208.6449 [hep-th]}}.

\bibitem{Alvarez-Gaume:2016vff}
L.~Alvarez-Gaume, O.~Loukas, D.~Orlando, and S.~Reffert, ``{Compensating strong
  coupling with large charge},''
\href{http://arxiv.org/abs/1610.04495}{{\tt arXiv:1610.04495 [hep-th]}}.

\bibitem{Ivanov:1981wn}
E.~A. Ivanov and J.~Niederle, ``{Gauge Formulation of Gravitation Theories. 1.
  The Poincare, De Sitter and Conformal Cases},''
\href{http://dx.doi.org/10.1103/PhysRevD.25.976}{{\em Phys. Rev.} {\bf D25}
  (1982)  976}.

\bibitem{Delacretaz:2014oxa}
L.~V. Delacretaz, S.~Endlich, A.~Monin, R.~Penco, and F.~Riva,
  ``{(Re-)Inventing the Relativistic Wheel: Gravity, Cosets, and Spinning
  Objects},'' \href{http://dx.doi.org/10.1007/JHEP11(2014)008}{{\em JHEP} {\bf
  11} (2014)  008},
\href{http://arxiv.org/abs/1405.7384}{{\tt arXiv:1405.7384 [hep-th]}}.

\bibitem{Karananas:2015eha}
G.~K. Karananas and A.~Monin, ``{Weyl and Ricci gauging from the coset
  construction},'' \href{http://dx.doi.org/10.1103/PhysRevD.93.064013}{{\em
  Phys. Rev.} {\bf D93} (2016) no.~6, 064013},
\href{http://arxiv.org/abs/1510.07589}{{\tt arXiv:1510.07589 [hep-th]}}.

\bibitem{Karananas:2016hrm}
G.~K. Karananas and A.~Monin, ``{Gauging nonrelativistic field theories using
  the coset construction},''
  \href{http://dx.doi.org/10.1103/PhysRevD.93.064069}{{\em Phys. Rev.} {\bf
  D93} (2016)  064069},
\href{http://arxiv.org/abs/1601.03046}{{\tt arXiv:1601.03046 [hep-th]}}.

\bibitem{Volkov:1973vd}
D.~V. Volkov, ``{Phenomenological Lagrangians},''
{\em Fiz. Elem. Chast. Atom. Yadra} {\bf 4} (1973)  3--41.

\bibitem{Ivanov:1975zq}
E.~A. Ivanov and V.~I. Ogievetsky, ``{The Inverse Higgs Phenomenon in Nonlinear
  Realizations},''
\href{http://dx.doi.org/10.1007/BF01028947}{{\em Teor. Mat. Fiz.} {\bf 25}
  (1975)  164--177}.

\bibitem{Low:2001bw}
I.~Low and A.~V. Manohar, ``{Spontaneously broken space-time symmetries and
  Goldstone's theorem},''
  \href{http://dx.doi.org/10.1103/PhysRevLett.88.101602}{{\em Phys. Rev. Lett.}
  {\bf 88} (2002)  101602},
\href{http://arxiv.org/abs/hep-th/0110285}{{\tt arXiv:hep-th/0110285
  [hep-th]}}.

\bibitem{Manohar:1983md}
A.~Manohar and H.~Georgi, ``{Chiral Quarks and the Nonrelativistic Quark
  Model},''
\href{http://dx.doi.org/10.1016/0550-3213(84)90231-1}{{\em Nucl. Phys.} {\bf
  B234} (1984)  189--212}.

\bibitem{Georgi:1992dw}
H.~Georgi, ``{Generalized dimensional analysis},''
  \href{http://dx.doi.org/10.1016/0370-2693(93)91728-6}{{\em Phys. Lett.} {\bf
  B298} (1993)  187--189},
\href{http://arxiv.org/abs/hep-ph/9207278}{{\tt arXiv:hep-ph/9207278
  [hep-ph]}}.

\bibitem{Cohen:1997rt}
A.~G. Cohen, D.~B. Kaplan, and A.~E. Nelson, ``{Counting 4 pis in strongly
  coupled supersymmetry},''
  \href{http://dx.doi.org/10.1016/S0370-2693(97)00995-7}{{\em Phys. Lett.} {\bf
  B412} (1997)  301--308},
\href{http://arxiv.org/abs/hep-ph/9706275}{{\tt arXiv:hep-ph/9706275
  [hep-ph]}}.

\bibitem{Nicolis:2013sga}
A.~Nicolis, R.~Penco, F.~Piazza, and R.~A. Rosen, ``{More on gapped Goldstones
  at finite density: More gapped Goldstones},''
  \href{http://dx.doi.org/10.1007/JHEP11(2013)055}{{\em JHEP} {\bf 11} (2013)
  055},
\href{http://arxiv.org/abs/1306.1240}{{\tt arXiv:1306.1240 [hep-th]}}.

\bibitem{Rychkov:2016iqz}
S.~Rychkov, ``{EPFL Lectures on Conformal Field Theory in D>= 3 Dimensions},''
\href{http://arxiv.org/abs/1601.05000}{{\tt arXiv:1601.05000 [hep-th]}}.

\bibitem{Bilal:2013iva}
A.~Bilal and F.~Ferrari, ``{Multi-Loop Zeta Function Regularization and
  Spectral Cutoff in Curved Spacetime},''
  \href{http://dx.doi.org/10.1016/j.nuclphysb.2013.10.003}{{\em Nucl. Phys.}
  {\bf B877} (2013)  956--1027},
\href{http://arxiv.org/abs/1307.1689}{{\tt arXiv:1307.1689 [hep-th]}}.

\bibitem{Assel:2015nca}
B.~Assel, D.~Cassani, L.~Di~Pietro, Z.~Komargodski, J.~Lorenzen, and
  D.~Martelli, ``{The Casimir Energy in Curved Space and its Supersymmetric
  Counterpart},'' \href{http://dx.doi.org/10.1007/JHEP07(2015)043}{{\em JHEP}
  {\bf 07} (2015)  043},
\href{http://arxiv.org/abs/1503.05537}{{\tt arXiv:1503.05537 [hep-th]}}.

\bibitem{Monin:2016bwf}
A.~Monin, ``{Partition function on spheres: How to use zeta function
  regularization},'' \href{http://dx.doi.org/10.1103/PhysRevD.94.085013}{{\em
  Phys. Rev.} {\bf D94} (2016) no.~8, 085013},
\href{http://arxiv.org/abs/1607.06493}{{\tt arXiv:1607.06493 [hep-th]}}.

\end{thebibliography}\endgroup

\end{document}